\documentclass[preprint,aps,notitlepage,nofootinbib,superscriptaddress]{revtex4}
\usepackage{graphicx}
\usepackage{color}
\usepackage{bm}
\usepackage{multirow}
\usepackage{amsmath}
\usepackage{amsfonts}

\begin{document}

\newcommand{\YITP}{
  Center for Gravitational Physics, Yukawa Institute for Theoretical Physics, 
  Kyoto University, 
  Kyoto 606-8502, Japan
}
\newcommand{\Osaka}{
  Department of Physics, Osaka University,
  Toyonaka 560-0043, Japan
}
\newcommand{\Edinburgh}{
School of Physics and Astronomy, The University of Edinburgh, Edinburgh EH9 3JZ, United Kingdom 
}
\newcommand{\KEK}{
  High Energy Accelerator Research Organization (KEK), 
  Tsukuba 305-0801, Japan
}
\newcommand{\Sokendai}{
  School of High Energy Accelerator Science,
  The Graduate University for Advanced Studies (Sokendai), 
  Tsukuba 305-0801, Japan
}
\newcommand{\RIKENCCS}{
RIKEN Center for Computational Science,
7-1-26 Minatojima-minami-machi, Chuo-ku, Kobe, Hyogo 650-0047, Japan
}
\newcommand{\JAEA}{
Advanced Science Research Center, Japan Atomic Energy Agency (JAEA), Tokai 319-1195, Japan
}
\preprint{YITP-20-126, OU-HET-905, KEK-CP-0378}

\title{
Study of the axial $U(1)$ anomaly at high temperature with lattice chiral fermions   
}
\author{S.~Aoki}
\affiliation{\YITP}
\author{Y.~Aoki}
\affiliation{\RIKENCCS}
\author{G.~Cossu}
\affiliation{\Edinburgh}
\author{H.~Fukaya}
\affiliation{\Osaka}
\author{S.~Hashimoto}
\affiliation{\KEK}
\affiliation{\Sokendai}
\author{T.~Kaneko}
\affiliation{\KEK}
\affiliation{\Sokendai}
\author{C.~Rohrhofer}
\affiliation{\Osaka}
\author{K.~Suzuki}
\affiliation{\JAEA}

\collaboration{JLQCD collaboration}
\noaffiliation

\begin{abstract}
  We investigate the axial $U(1)$ anomaly of two-flavor QCD
  at temperatures 190--330 MeV.
  In order to preserve precise chiral symmetry on the lattice, we employ
  the M\"obius domain-wall fermion action as well as
  overlap fermion action implemented with a stochastic reweighting technique.
Compared to our previous studies, we reduce the lattice spacing
to 0.07 fm, simulate larger multiple volumes to estimate finite size effect,
and take more than four quark mass points,
including one below physical point to investigate the chiral limit.
We measure the topological susceptibility, axial $U(1)$ susceptibility, and
examine the degeneracy of $U(1)$ partners in meson/baryon correlators.
All the data above the critical temperature
indicate that the axial $U(1)$ violation is consistent with zero within statistical errors.
The quark mass dependence suggests disappearance of the  $U(1)$ anomaly
at a rate comparable to that of the $SU(2)_L\times SU(2)_R$ symmetry breaking.
\end{abstract}
\maketitle
\section{Introduction}

The two-flavor QCD Lagrangian in the massless limit has
a global $SU(2)_L\times SU(2)_R \times U(1)_V \times U(1)_A$ symmetry.
It is widely believed that its $SU(2)_L\times SU(2)_R$ part is
spontaneously broken to $SU(2)_V$ at low temperatures but is restored
above some critical temperature,
which is called the chiral phase transition.
On the other hand, the axial $U(1)_A$ part is broken by the chiral anomaly.
Since the anomaly refers to symmetry breaking at the cut-off scale
where the theory is defined,
and the anomalous Ward-Takahashi identity holds at any temperature \cite{Itoyama:1982up},
it is natural to assume that the anomaly survives the chiral phase transition
and the physics of the early universe is not $U(1)_A$ symmetric.

However, there is a counter argument to this naive picture.
In fact,  the $U(1)_A$ anomaly is connected to the topology of
the gauge field, which is sensitive to the low energy dynamics, and
low-lying modes of the Dirac operator
affect the strength of the $U(1)_A$ violation.
In particular, if the Dirac spectrum has a gap at the lowest end of the spectrum,
the $U(1)_A$ anomaly becomes invisible in two-point mesonic correlation functions
in the chiral limit \cite{Cohen:1996ng, Cohen:1997hz}.
Since the $SU(2)_L\times SU(2)_R$ symmetry is also related to
the Dirac spectrum through the Banks-Casher relation \cite{Banks:1979yr},
the restoration of $SU(2)_L\times SU(2)_R$ at the chiral transition
may also affect the $U(1)_A$ anomaly. Indeed, it was argued
that the $U(1)_A$ anomaly can completely disappear
in correlation functions of scalar and pseudoscalar operators \cite{Aoki:2012yj}.

If the effect of the $U(1)_A$ anomaly is negligible at and above
the critical temperature,
it may have an impact on our understanding of the
phase diagram of QCD \cite{Pisarski:1983ms}.
In the standard effective theory analysis of the
chiral phase transition, we expect only the pion and
its chiral-partner scalar particle to be the light degrees of freedom
that govern the low-energy dynamics of QCD
near the phase transition.
When the $U(1)_A$ symmetry is effectively restored,
the $\eta$, isospin singlet pseudoscalar, and its chiral partner may
play a nontrivial role \cite{Horvatic:2018ztu, Bass:2018xmz, Nicola:2019ohb}.
Since the effective potential can have more complicated structure as the degrees of freedom
increase, it was argued  that the
chiral phase transition likely becomes first order
when $U(1)_A$ symmetry is recovered
(we refer the readers to \cite{Yonekura:2019vyz} for a different aspect of
the first-order scenario from 't Hooft anomaly matching),
though other scenarios are theoretically possible
\cite{Pelissetto:2013hqa, Nakayama:2014sba, Kanazawa:2014cua, Sato:2014axa, Nakayama:2016jhq, Alexandru:2019gdm}.

How much the $U(1)_A$ anomaly contributes to the dynamics has a significant
importance on cosmology.
The topological susceptibility is related to
the mass and decay constant of the QCD axion,
which is a candidate of the dark matter.
Its temperature dependence influences 
the relic abundance of the axion
\cite{Berkowitz:2015aua,Kitano:2015fla,Bonati:2015vqz,
  Petreczky:2016vrs,Borsanyi:2016ksw, Burger:2018fvb,Lombardo:2020bvn}.

For the nontrivial question of how much the $U(1)_A$ anomaly remains
and affects the chiral phase transition,
only lattice QCD can give a quantitative answer.
Good control of chiral symmetry on the lattice
\cite{Bazavov:2012qja, Cossu:2013uua, Buchoff:2013nra, Chiu:2013wwa}
is necessary in order to
precisely discriminate between the lattice artifact and 
the physical signal that survives in the continuum limit.
Our previous studies \cite{Cossu:2015kfa, Cossu:2016scb, Tomiya:2016jwr}
demonstrated that the signal of topological susceptibility
is sensitive to the violation of chiral symmetry at high temperature,
and the reweighting of the M\"obius domain-wall fermion determinant to that of the overlap fermion
is essential when the lattice spacing is coarse $a\gtrsim 0.1$ fm.
We also found that the use of the overlap fermion
only in the valence sector \cite{Dick:2015twa, Sharma:2018syt,Mazur:2018pjw}
makes the situation worse,
since the lattice artifact due to the mixed action, which is unphysical, is strongly enhanced.

After removing the lattice artifact due to the violation
of the Ginsparg-Wilson relation at high temperature,
we observed that chiral limit of the $U(1)_A$ susceptibility
is consistent with zero \cite{Tomiya:2016jwr}.
The disappearance of the $U(1)_A$ anomaly (at around 1.2 $T_c$)
was also reported by other groups
simulating non-chiral fermions \cite{Brandt:2016daq, Ishikawa:2017nwl}\footnote{
  In \cite{Brandt:2019ksy}, they gave a preliminary results showing that
  the anomaly studied in \cite{Brandt:2016daq} looks enhanced as the volume size increases.
}.
In \cite{Bazavov:2019www}, it was found that the
$U(1)_A$ symmetry shows up at 1.3 $T_c$ but not around $T_c$
\footnote{
  A recent work \cite{1826587} reported that $U(1)_A$ symmetry is still broken at 1.6 $T_c$
  using  highly improved staggered quarks with lattice spacings 0.06--0.12 fm.
  }.

In this work, we reinforce the conclusion of \cite{Tomiya:2016jwr}
by 1) reducing the lattice spacing to an extent where
we observe consistency between the overlap and M\"obius domain-wall fermions,
2) simulating different lattice volume sizes in a range $1.8 \le L \le 3.6$ fm,
and 3) simulating more quark mass points, including one below the physical point,
to investigate the chiral limit.
In order to study possible artifact due to topology freezing,
we also apply the reweighting from a larger quark mass, where topology tunneling is frequent,
down to the mass where the topological susceptibility is consistent with zero.
We measure the topological susceptibility, axial $U(1)$ susceptibility, meson correlators,
and baryon correlators.
Some of the results were already reported in our contributions
to the conference proceedings
\cite{Aoki:2017xux, Suzuki:2017ifu, Fukaya:2017wfq, Suzuki:2018vbe,
  Suzuki:2019vzy, Rohrhofer:2019yko, Suzuki:2020rla}.

All the data at temperatures above 190 MeV 
show that the axial $U(1)$ anomaly is consistent with zero, within statistical errors.
Its quark mass dependence indicates that the disappearance of the  $U(1)$ anomaly
is at a rate comparable to that of the $SU(2)_L\times SU(2)_R$ symmetry.
At higher temperature, we have also observed
a further enhancement of symmetry
\cite{Rohrhofer:2017grg, Glozman:2017dfd, Lang:2018vuu, Rohrhofer:2019qwq}.

The rest of the paper is organized as follows.
We describe our lattice setup
and how to implement the chiral fermions in the simulations in Sec.~\ref{sec:setup}.
In Sec.~\ref{sec:results}, the numerical results for the
Dirac spectrum, topological susceptibility,
axial $U(1)$ susceptibility, and meson/baryon screening masses
are presented.
The conclusion is given in Sec.~\ref{eq:conclusion}.

\section{Lattice setup}
\label{sec:setup}

The setup of our simulations is basically the same
as our previous study \cite{Tomiya:2016jwr}, except for the choice of
parameters
(larger lattice sizes up to $3.6$ fm,
and smaller lattice spacing $a\sim 0.074$ fm).
Our naive estimate for the critical temperature of the chiral phase transition
in \cite{Tomiya:2016jwr}
is $T_c\sim 175$ MeV, which was obtained from the Polyakov loop\footnote{
  Our estimate for the critical temperature on coarse and small lattices
  was not very accurate
  whereas the scale setting via the Wilson flow is precise.
  See below for the details.
  }.
Below, we summarize the essential part.

In the hybrid-Monte-Carlo (HMC) simulations\footnote{
  Numerical works are done with the QCD software package IroIro$++$ \cite{Cossu:2013ola}, Grid \cite{Boyle:2015tjk}
  and Bridge++\cite{Ueda:2014rya}.
},
we employ the tree-level improved Symanzik gauge action \cite{Luscher:1985zq} 
for the link variables and the domain-wall fermion \cite{Kaplan:1992bt}
with an improvement
(the M\"obius domain-wall fermions \cite{Brower:2005qw, Brower:2012vk})
for the quark fields.
Here we set the size of the fifth direction as $L_s=16$.
The M\"obius kernel is taken as 
\begin{equation}
H_M = \gamma_5\frac{2D_W}{2+D_W},
\end{equation}
where $D_W$ is the standard Wilson-Dirac operator
with a large negative mass $-1/a$.
In the following, we omit $a$, when there is no risk of confusion.
The numbers without physical unit, {e.g.} MeV, are in the lattice unit.
The fermion determinant
thus obtained
corresponds to a four-dimensional effective operator, 
\begin{eqnarray}
  \label{eq:4DDW}
D_{\rm DW}^{\rm 4D}(m) = \frac{1+m}{2}+\frac{1-m}{2}\gamma_5 \tanh(L_s\tanh^{-1}(H_M)).
\end{eqnarray}
Note that if we take the $L_s=\infty$ limit, this operator
converges to the overlap-Dirac operator \cite{Neuberger:1997fp}
with a kernel operator $H_M$.
The stout smearing \cite{Morningstar:2003gk} is applied three times
with the smoothing parameter $\rho=0.1$
for the link variables in the Dirac operator.

The residual mass for our main runs at $\beta=4.30$ is 0.14(6) MeV, 
which may look small enough for the measurements in this work. 
However, as we reported in our previous studies,
the chiral symmetry breaking due to lattice artifacts
is enhanced at high temperature, in particular
for the axial $U(1)$ susceptibility.
Therefore, we employ an improved Dirac operator,
exactly treating the near-zero modes of $H_M$ to compute the
sign function.
With the 
near-zero modes whose absolute
value is less than 0.24 ($\sim 630$ MeV) treated exactly, we find that
the residual mass becomes negligible $\sim 10^{-5}$ or $\sim$ 0.01 MeV.
Although it is different from the original definition in \cite{Neuberger:1997fp},
we call this improved Dirac operator the ``overlap'' Dirac operator $D_{ov}(m)$.
Note that our overlap operator has a finite $L_s$, {\it i.e.} $L_s=16$.
As we also found in our previous study that
the lattice artifact from mixed action is
large even though the overlap and M\"obius domain-wall fermion actions
are very similar to each other, we reweight the gauge configurations of 
the M\"obius domain-wall fermion determinant
by that of the overlap fermion.
For details of this OV/MDW reweighting, see \cite{Tomiya:2016jwr}. 
As we will see  below, at $\beta=4.30$ the results with
the overlap fermion and those with
M\"obius domain-wall fermion are consistent with each other,
except for the $U(1)$ susceptibility at $T\leq 260$ MeV.
For the meson and baryon correlators, we use the M\"obius domain-wall fermion
without reweighting.

The lattice spacing is estimated from the Wilson flow with a
reference flow time $t_0=(0.1539\;\mbox{fm})^2$
determined in \cite{Sommer:2014mea}. 
For our main runs at $\beta=4.30$, the physical point of
the bare quark mass is estimated
as $m=0.0014(2)$, which is slightly above our lightest quark mass.
For this estimate, we used a leading-order chiral perturbation formula,
with an input of simulated pion mass $m_\pi =0.135(8)$ (in the lattice units)
determined from a zero temperature study at $m=0.01$.
The choice of the simulation parameters enables us to interpolate the results to the physical point.

The simulation parameters are summarized in Tab.~\ref{tab:prms}.
Compared to the previous work \cite{Tomiya:2016jwr}, we reduce the lattice spacing from $\sim 0.1$ fm to 0.074 fm.
At this value of lattice spacing, we change
temperatures by varying the temporal extent $L_t=8,10,12,14$,
which correspond to temperatures $T=$ 190--330 MeV.
In order to estimate the finite volume effects,
we simulate with four different volume sizes $L^3$ with $L=24,32,40$ and 48 at $T=220$ MeV.
We also increase the statistics at $\beta=4.24$ ensembles continued from \cite{Tomiya:2016jwr}
to check the consistency between different lattice spacings.

For each ensemble, we simulated more than 20000 trajectories from which
we carry out measurements on the configurations separated by 100 trajectories.
We then bin the data in every 1000 trajectories,
which is longer than autocorrelation lengths we observe.
When we use the OV/MDW reweighting, we lose some amount of statistics due to its noise.
An estimate for the effective number of statistics $N_{\rm rew}$ in the table is defined as
\begin{eqnarray}
  N_{\rm rew} =\frac{ \langle R\rangle}{R_{\rm max}},
\end{eqnarray}
where $R$ is the reweighting factor or  the ratio of the determinant
and $R_{\rm max}$ is its maximal value in the ensemble.
As discussed in \cite{Tomiya:2016jwr}, $R$ has a negative correlation
with the axial $U(1)$ observables, and the statistical error estimated by
the jackknife method are smaller than expected from $1/\sqrt{N_{\rm rew}}$.
For the ensembles with small number of $N_{\rm rew}\sim 10$,
it is important to check the consistency with the M\"obius domain-wall
results without reweighting.

\renewcommand{\arraystretch}{0.5}
\begin{table}[tbp]
  \centering
  \begin{tabular}{ccccccccc}
    \hline\hline
    $\beta$ & $a$(fm) & $L^3\times L_t$ & $T$(MeV) & $L$(fm) & $m$ & \#trj. & $N_{\rm rew}$ & comments \\
    \hline
    4.24 & 0.084 & $32^3\times 12$ & 195 & 2.7 &  0.0025 & 21200 & 10(2) \\
  &  &  &  &  & 0.005 & 20000 & 10(1) & continuation from  \cite{Tomiya:2016jwr}\\
 &  &  &  &  & 0.01 & 25300 & 7(1) \\
 \hline
4.30 & 0.074 & $32^3\times 14$ & 190 & 2.4 & 0.001 & 13900 & 38(2) \\
 &  &  &  &  & 0.0025 & 16600 & 8(1) \\
 &  &  &  &  & 0.00375 & 12500 & 10(1) \\
 &  &  &  &  & 0.005 & 10600 & 9(1) \\
\cline{3-9}
 & & $24^3\times 12$ & 220 & 1.8 & 0.001 & 31900 & 50(1) \\
 &  &  &  &  & 0.0025 & 33400 & 49(1) \\
 &  &  &  &  & 0.00375 & 34600 & 17(1) \\
 &  &  &  &  & 0.005 & 36000 & 16(1) \\
 &  &  &  &  & 0.01 & 35900 & 47(2) \\
\cline{3-9}
&  & $32^3\times 12$ & 220 & 2.4 & 0.001 & 26500 & 32(1) \\
 &  &  &  &  & 0.0025 & 26660 & 57(2) \\
 &  &      &  &  & 0.00375 & 26420 & 28(2) \\
 &  &      &  &  & 0.005 & 18560 & 30(1) \\
 &  &      &  &  & 0.01 & 31000 & 93(2) \\
 \cline{3-9}
&  & $40^3\times 12$ & 220 & 3.0 & 0.005 & 28100 & 13(1) \\
 &  &  &  &  & 0.01 & 27300 & 39(2) \\
 \cline{3-9}
&  & $48^3\times 12$ & 220 & 3.6 & 0.001 & 11200 & 4(1) \\
 &  &  &  &  & 0.0025 & 11300 & 8(1) \\
 &  &      &  &  & 0.00375 & 12800 & 10(2) \\
 &  &      &  &  & 0.005 & 10900 & 2(1) \\
 \cline{3-9}
&  & $32^3\times 10$ & 260 & 2.4 & 
0.005 & 12780 & 25(1) & $m_r=0.003,0.004$\\
 &  &  &  &  & 0.008 & 20050 & 19(1) \\
 &  &  &  &  & 0.01 & 29000 & 58(2) & $m_r=0.006,0.007,0.008,0.009$  \\
 &  &  &  &  & 0.015 & 12000 & 20(1) \\
 \cline{3-9}
 &  & $32^3\times 8$ & 330 & 2.4
 & 0.001 & 26100 & 42(2) \\
 &&&& & 0.005 & 31700 & 28(2) \\
&&&& & 0.01 & 24500 & 37(5) \\
 &  &  &  &  & 0.015 & 30500 & 61(2) \\
 &  &  &  &  & 0.02 & 19100 & 39(2) & $m_r=0.0125,0.015,0.0175$\\
 &  &  &  &  & 0.04 & 5000 & 12(1) \\
  \cline{3-9}
 &  & $48^3\times 8$ & 330 & 3.6
  & 0.01 & 9000 & 19(2) \\
 &  &  &  &  & 0.015 & 13950 & 17(2) \\
\hline\hline
  \end{tabular}
  \caption{Simulation parameters. The values of $m_r$ are
    taken for the mass reweighting on that particular ensemble.
    $N_{\rm rew}$ denotes the effective number of
    configurations after the OV/MDW reweighting (see the main text).
    The $\beta=4.24$ runs are continuation from \cite{Tomiya:2016jwr} to check the consistency.}
  \label{tab:prms}
\end{table}
\clearpage
\section{Numerical results}
\label{sec:results}

In this section, we show our numerical results
on the axial $U(1)$ anomaly.
For meson and baryon correlators,
we also present the tests
of the $SU(2)_L\times SU(2)_R$ symmetry for comparison.

\subsection{Dirac spectrum}

The spectral density of the Dirac operator
\begin{eqnarray}
\rho(\lambda) = \frac{1}{V}\sum_i \langle \delta(\lambda-\lambda_i)\rangle,
\end{eqnarray}
in volume $V$ is used as a probe of chiral symmetry breaking.
Here, $\lambda_i$ denotes the $i$-th
eigenvalue of the Dirac operator on a given gauge configuration
and $\langle \cdots \rangle$
is the gauge ensemble average.
In the chiral limit after taking the thermodynamical limit,
we obtain the Banks-Casher relation \cite{Banks:1979yr},
which relates the chiral condensate $\langle \bar{q}q \rangle$
to the spectrum at $\lambda=0$:
\begin{eqnarray}
\langle \bar{q}q \rangle = \pi \rho(0).
\end{eqnarray}
We expect that $\rho(0)=0$
above the critical temperature of the chiral phase transition.

As the vacuum expectation value (vev) $\langle \bar{q}q \rangle$
also breaks the axial $U(1)$ symmetry,
the details of $\rho(\lambda)$ in the vicinity of zero is important in this work.
It has been shown that if the spectrum has a finite gap at $\lambda=0$,
the anomaly becomes invisible in mesonic two-point functions
\cite{Cohen:1996ng, Cohen:1997hz}.
In Ref.~\cite{Aoki:2012yj}, it is argued that
the $SU(2)_L \times SU(2)_R$ symmetry
restoration requires $\rho(\lambda)\sim \lambda^\alpha$ with
the power $\alpha>2$, which is sufficient to show
the absence of the axial $U(1)_A$ anomaly in
multi-point correlation functions of scalar and pseudoscalar operators.

We measure the eigenvalues/eigenfunctions
of the Hermitian four-dimensional effective
Dirac operator $H_{DW}(m)=\gamma_5D_{\rm DW}^{\rm 4D}(m)$
and those of the corresponding overlap-Dirac operator.
Each eigenvalue $\lambda_m$ is converted to
the one of massless Dirac operator by
$\lambda = \sqrt{\lambda_m^2-m^2}/\sqrt{1-m^2}$.
Forty lowest (in their absolute value)
modes are stored for gauge configurations separated by 100 trajectories.
They cover a range from zero to 300--500 MeV.

\begin{figure}[tbp]
  \centering
  \includegraphics[width=14cm]{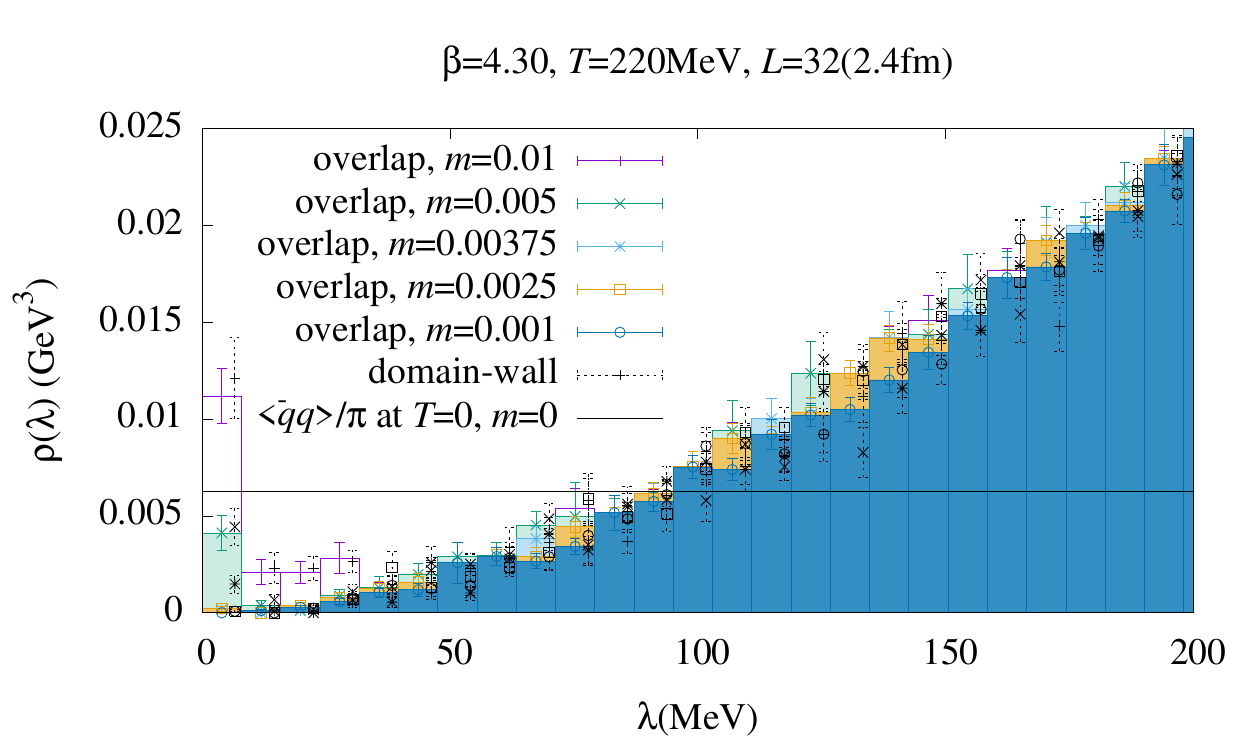}
  \caption{
    Dirac eigenvalue density at $T=220$ MeV.
    Solid symbols are the reweighted overlap results and
    dashed symbols are those of the M\"obius domain-wall.
    Horizontal line shows the chiral limit at $T=0$
    \cite{Cossu:2016eqs}.
  }
  \label{fig:Dirac220}
\end{figure}

In Fig.~\ref{fig:Dirac220}, we present the
Dirac eigenvalue density of the overlap-Dirac operator at $T=220$ MeV at $L=32$.
Here the OV/MDW reweighting is applied.
Compared to the solid line which represents the
chiral condensate at $T=0$ \cite{Cossu:2016eqs}, 
the low-modes are suppressed by an order of magnitude.
We observe a sharp peak near zero, which
rapidly disappears as quark mass decreases,
as expected from the $SU(2)_L\times SU(2)_R$ symmetry restoration.
Since we find no clear gap, a region of $\lambda$ where $\rho(\lambda)=0$,
we are not able to conclude if the
axial $U(1)$ symmetry is recovered or not from this observable only.

\begin{figure*}[tbp]
  \centering
  \includegraphics[width=14cm]{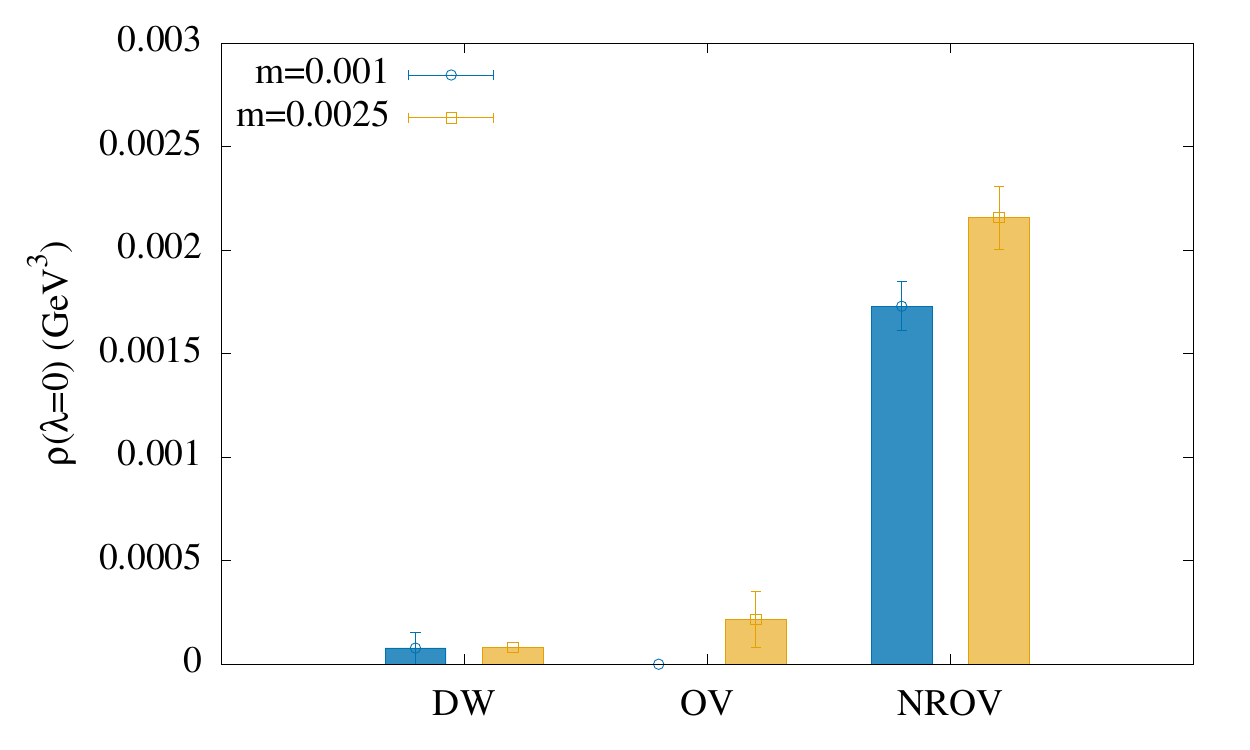}
  \caption{
    Comparison among  M\"obius domain-wall (MDW), reweighted overlap (OV),
    and non-reweighted overlap (NROV) data at
    the lowest bin with a bin size $\sim$ 10 MeV.
    A striking enhancement is seen for NROV, which
    is due to the lattice artifact in the mismatch of the valence and sea actions.
  }
  \label{fig:Dirac220sys}
\end{figure*}

We also compare these results
with the Dirac spectral density of the  M\"obius domain-wall fermion (dashed symbols) in Fig.~\ref{fig:Dirac220}.
The agreement is remarkable.
On the other hand, 
when we switch off the OV/MDW reweighting, which we call the non-reweighted overlap fermion setup,
we observe a remarkable  peak at the lowest bin, as presented in Fig.~\ref{fig:Dirac220sys}.
Similar peaks were reported in \cite{Dick:2015twa, Sharma:2018syt,Mazur:2018pjw},
with overlap fermion only in the valence sector.
Our data clearly show that these peaks are
due to the lattice artifact of the mixed action.

\begin{figure*}[tbp]
  \centering
  \includegraphics[width=14cm]{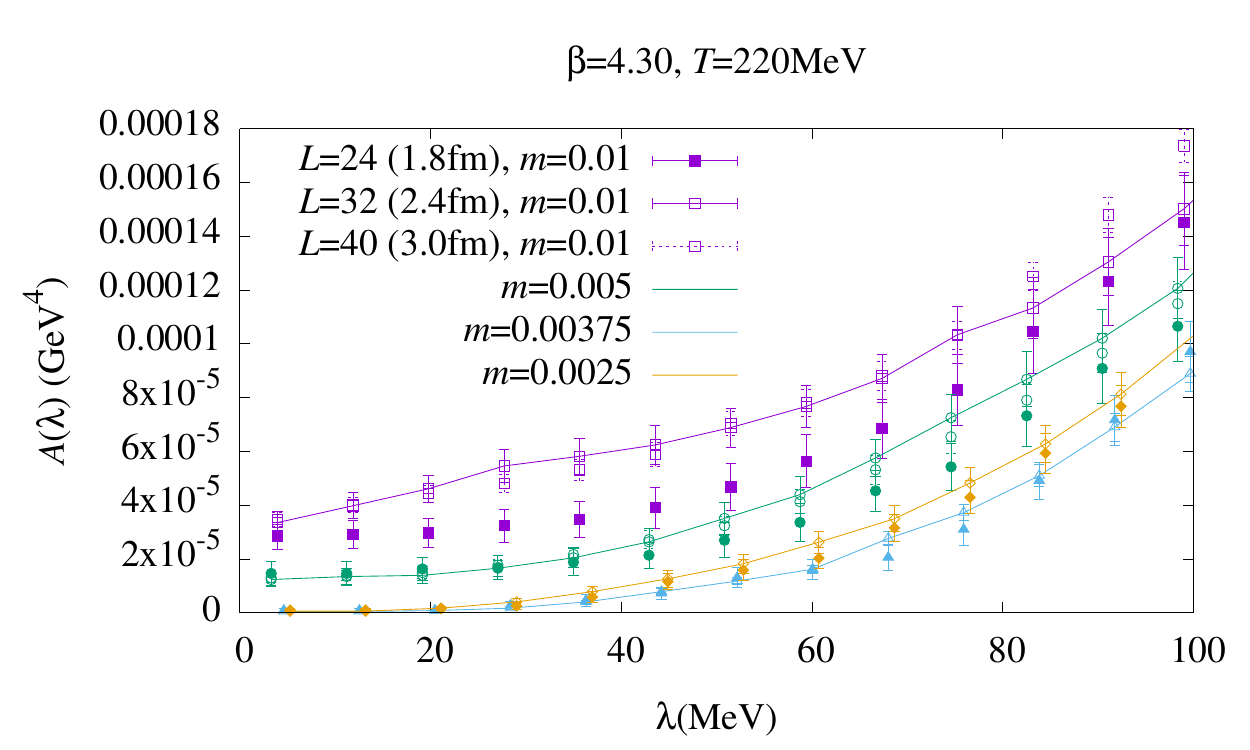}
  \caption{
    Accumulated Dirac histogram at three different volumes.
    Data at $m=0.01$ (square symbols), $0.005$ (circles), $0.00375$ (triangles) and $0.0025$ (diamonds)
    with $L=24$ (filled), $L=32$ (open with solid line) and
    $L=40$ (open dashed) are shown.
    Except for $m=0.01$ (squares), the data at different
    volume sizes are consistent.
  }
  \label{fig:Dirac220sys2}
\end{figure*}

In order to grasp possible systematic effects due to the finite lattice size,
we compare the accumulated histogram
\begin{eqnarray}
A(\lambda) = \int^\lambda_0 d\lambda' \rho(\lambda')
\end{eqnarray}
at three different volumes,
$L=24$ (1.8 fm), 32 (2.4 fm), 40 (3.0 fm) in Fig.~\ref{fig:Dirac220sys2}.
Except for $L=24$ at $m=0.01$,
whose aspect ratio $LT$ is $2$, which is smallest among the data sets,
no clear volume dependence is seen. 
The point $m=0.01$ is the heaviest mass in our simulation set,
which is expected to be least sensitive to the volume,
but the Dirac low-mode density is rather
high, and some remnants of spontaneous $SU(2)_L\times SU(2)_R$ breaking
and associated pseudo Nambu-Goldstone bosons may be responsible for
this volume dependence.

\begin{figure*}[tbp]
  \centering
  \includegraphics[width=8cm]{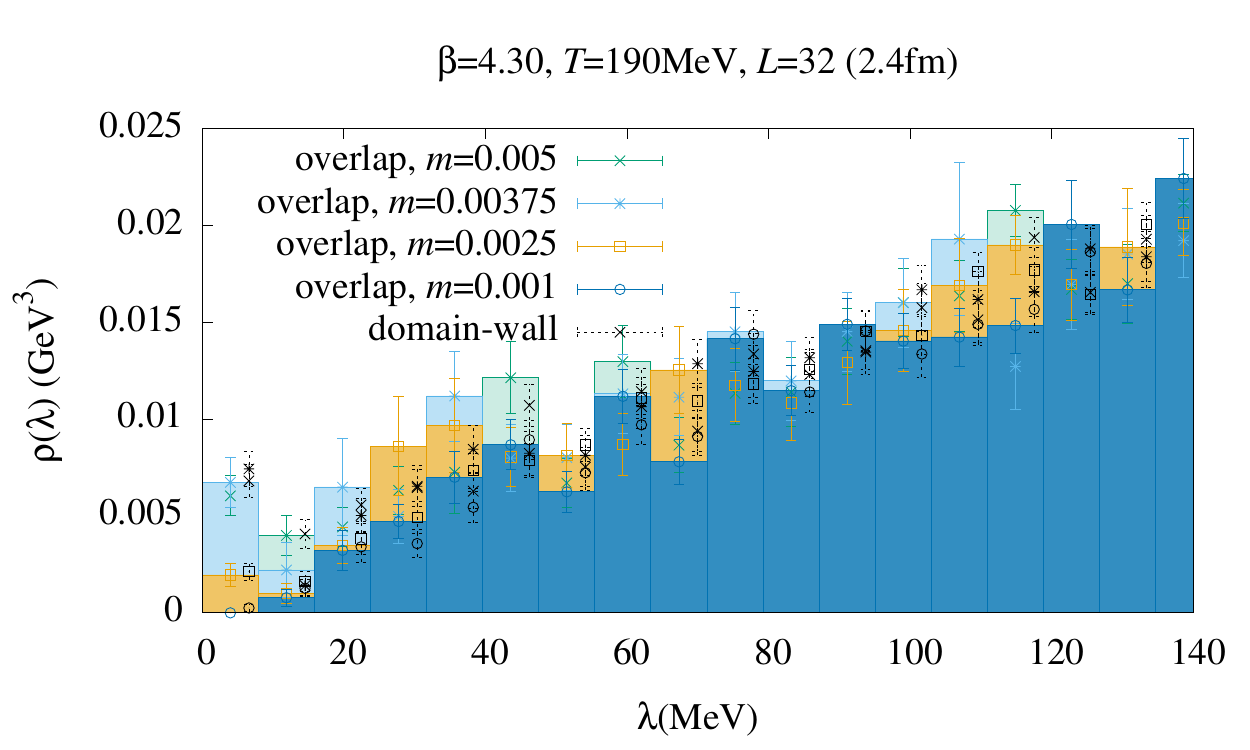}
  \includegraphics[width=8cm]{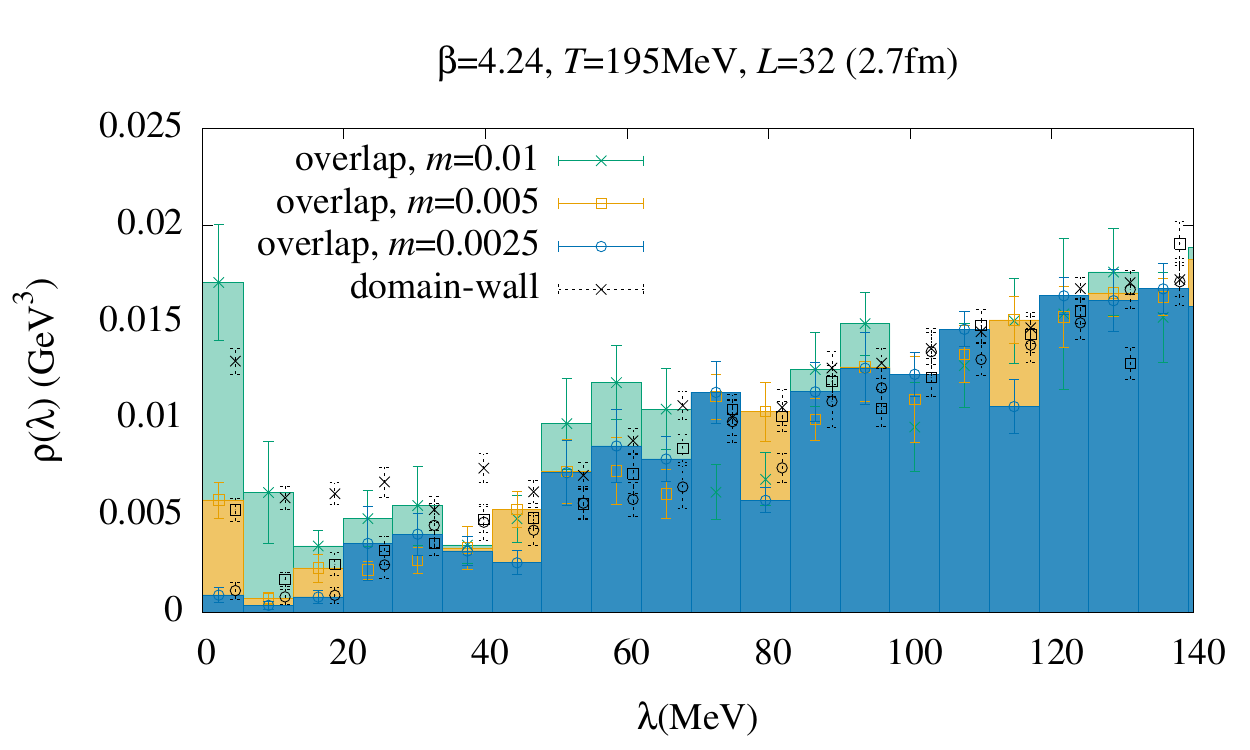}
  \includegraphics[width=8cm]{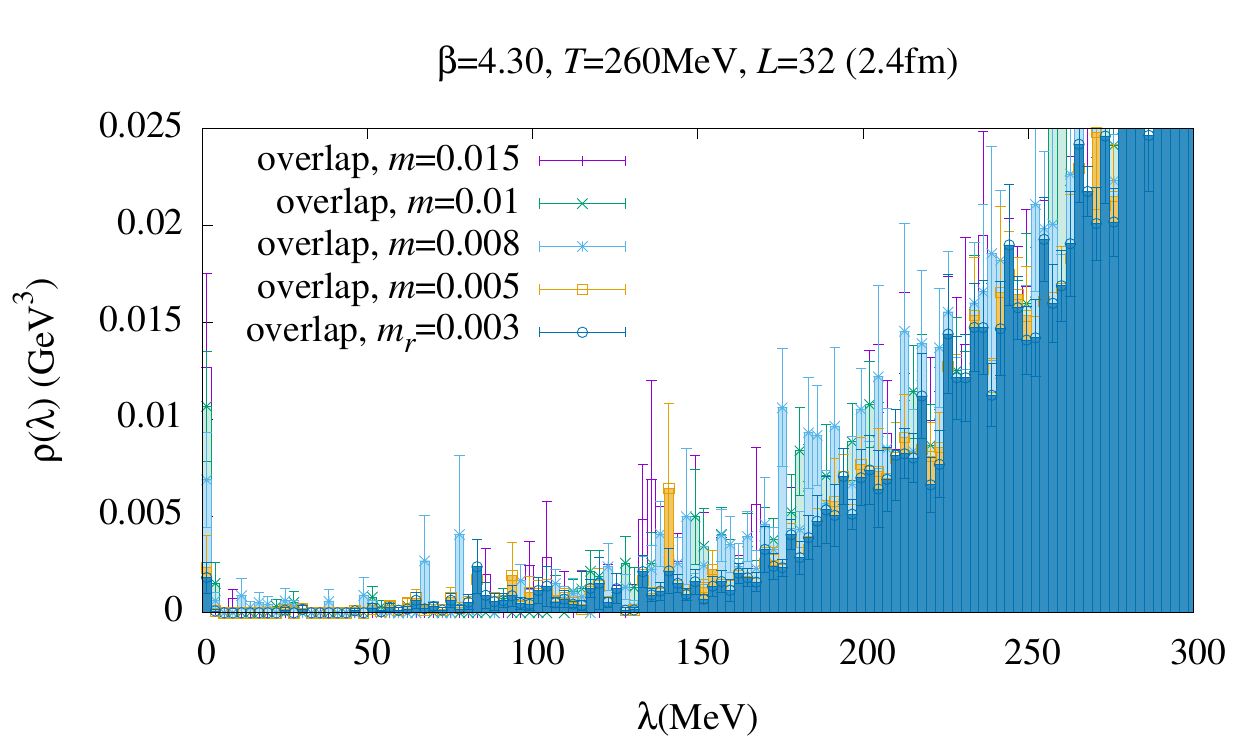}
  \includegraphics[width=8cm]{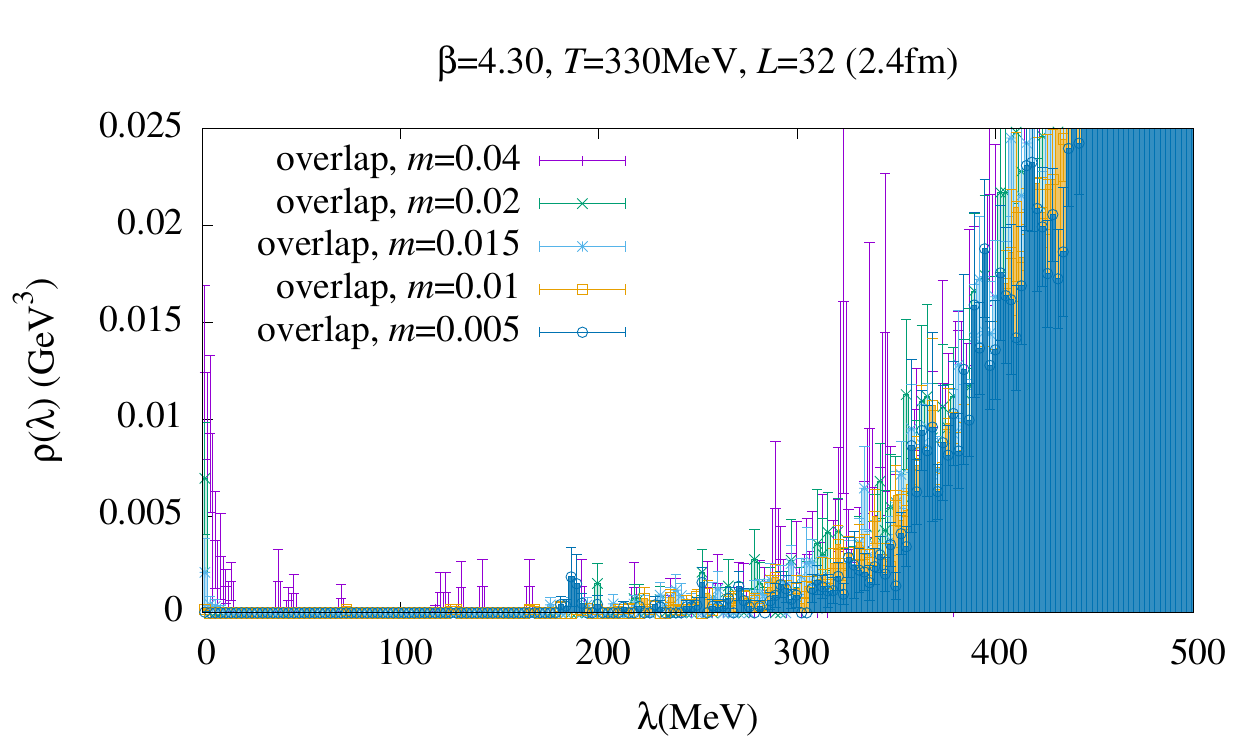}
  \caption{
    Dirac eigenvalue density at different temperatures.
  }
  \label{fig:DiracT}
\end{figure*}

We summarize the results at different temperatures in Fig.~\ref{fig:DiracT}.
The higher the temperature, the stronger the suppression of the low modes is.
We find a good consistency between the M\"obius domain-wall and
reweighted overlap results.
We also find that $\beta=4.30$ and $4.24$ results are consistent.
The quark mass dependence is not very strong except for
the lowest bin near $\lambda=0$.

\begin{figure}[tbp]
  \centering
  \includegraphics[width=14cm]{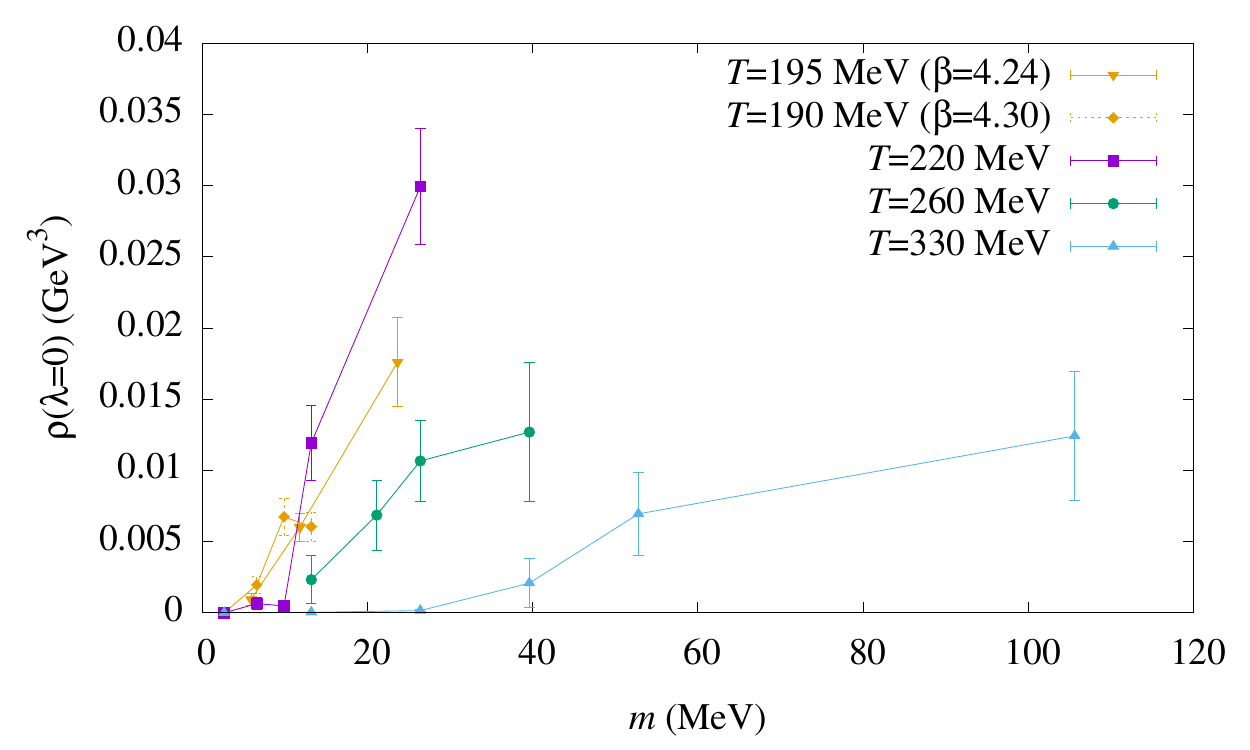}
  \caption{
    Quark mass dependence of the Dirac eigenvalue density at (near) zero
    with the bin size $\sim 10$ MeV.
    All the data look consistent with zero before reaching the chiral limit.
  }
  \label{fig:Dirac1stbin}
\end{figure}

Finally, the quark mass dependence of the eigenvalue density
of the reweighted overlap operator near $\lambda=0$
(with the bin size $\sim$ 10 MeV)
is presented in Fig.~\ref{fig:Dirac1stbin}.
Here both the chiral zero modes and non-chiral pair of
near zero-modes are included.
It is remarkable that the density of near-zero modes at different temperatures
show a steep decrease towards the massless limit
and becomes consistent with zero already at finite quark masses.
As will be shown below, this behavior of near zero modes
is strongly correlated with the signals of the axial $U(1)$ anomaly.


\clearpage
\subsection{Topological susceptibility}

In order to quantify the topological excitations of
gauge fields, we measure the topological charge $Q$ in  two different ways.
One is the index of the overlap-Dirac operator,
or the number of zero modes with positive chirality minus
that with negative chirality.
For this, we perform the OV/MDW reweighting
to avoid possible mixed action artifacts,
which is found to be quite significant.
The other is a gluonic definition, measured directly
on the  M\"obius domain-wall ensemble, using the clover-like construction of
the gauge field strength $F_{\mu\nu}$
after applying the Wilson flow on the gauge configuration
with a flow time $ta^2 = 5$.

\begin{figure}[tbp]
  \centering
  \includegraphics[width=14cm]{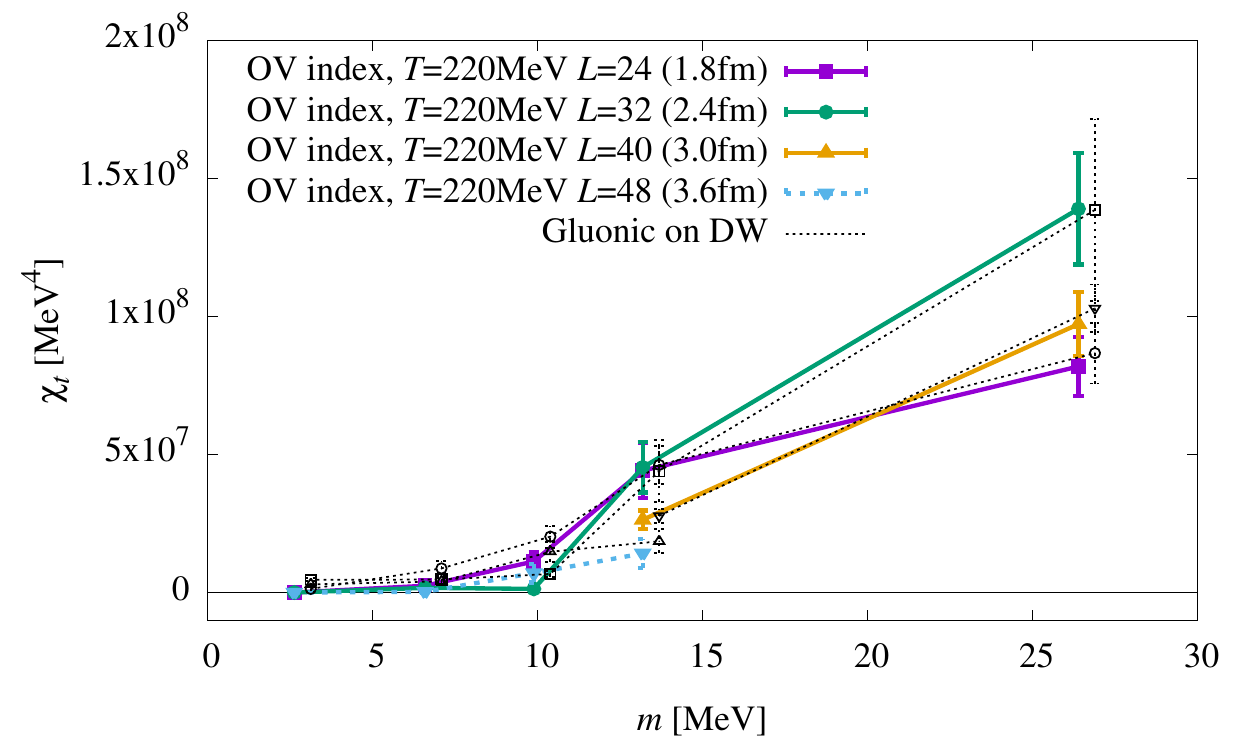}
  \caption{
    Topological susceptibility at $T=220$ MeV.
    The filled symbols with solid lines show the data for the index of
    the overlap-Dirac operator with the OV/MDW reweighting.
    Those with dashed lines are data at $L=48$ whose statistics may not be good enough
    after the reweighting.
    The dotted symbols are those for gluonic definition
    directly measured on the ensembles generated with the M\"obius domain-wall fermions.
    We confirm that both data with different volumes are consistent.
  }
  \label{fig:chit}
\end{figure}

The results for the topological susceptibility $\chi_t=\langle Q^2 \rangle/V$
obtained at $T=220$ MeV
are shown in Fig.~\ref{fig:chit}.
The filled symbols show the data for the index of
the overlap-Dirac operator with the OV/MDW reweighting,
while  the dotted symbols are those for gluonic definition
without reweighting.
Both are consistent with each other.
The systematics due to chiral symmetry violating lattice artifacts is
therefore under control.
We also find that there is no strong
volume dependence, except for the heaviest point with $L=1.8$ fm
where the aspect ratio is $LT=2$.

\begin{figure}[tbp]
  \centering
  \includegraphics[width=14cm]{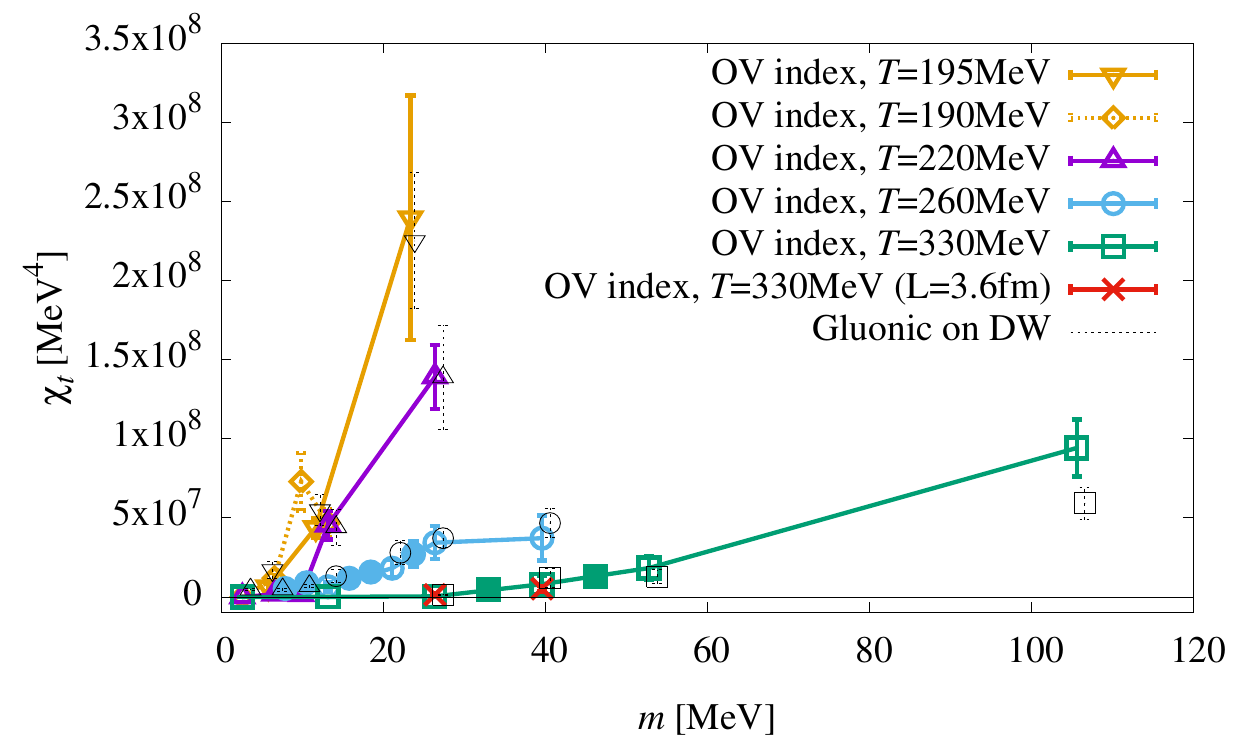}
  \caption{
    Topological susceptibility at different temperatures.
    The volume size is fixed to $L=32$ (2.4 fm), except for
    data at $T=195$ MeV ($\beta=4.24$) (2.7 fm) and those at $T=330$ MeV (3.6 fm)
    denoted by cross symbols.
    Filled symbols are obtained with reweighting
    from the ensemble at a higher mass point shown by open symbols.
  }
  \label{fig:chitT}
  \centering
  \includegraphics[width=14cm]{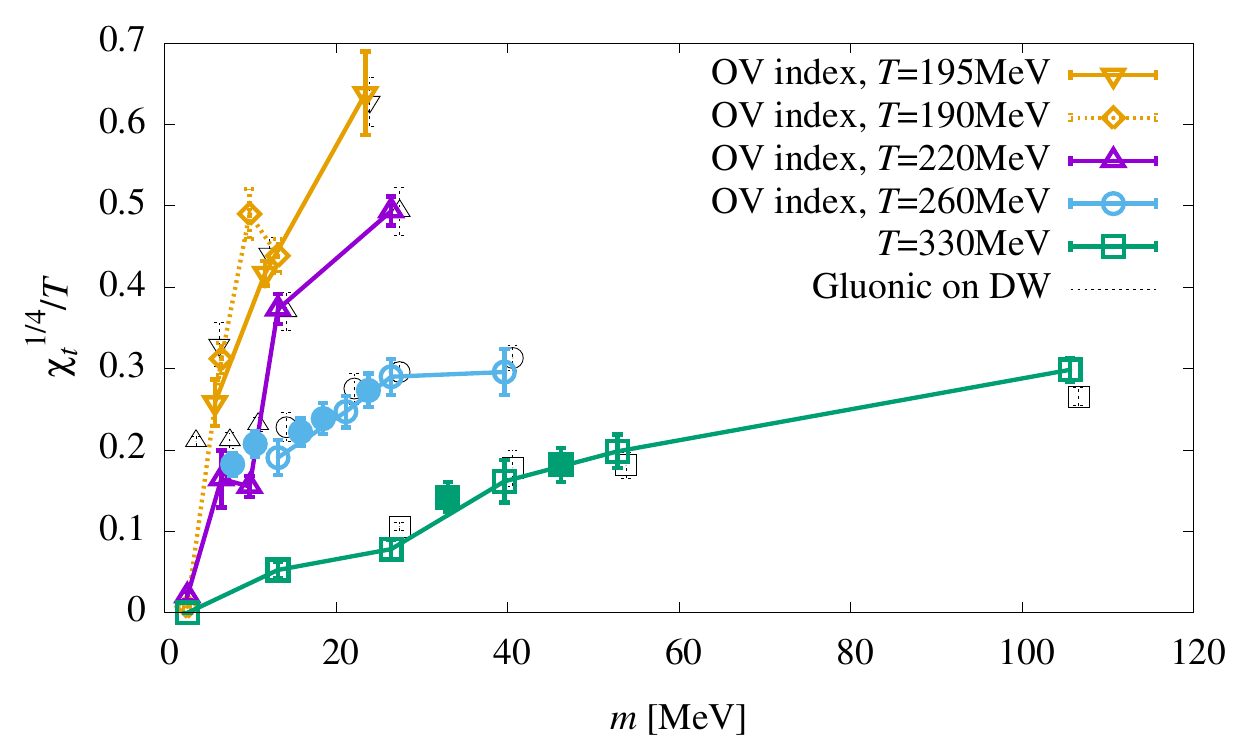}
  \caption{
    The same as Fig.~\ref{fig:chitT} but the
    fourth root is taken and normalized by $T$.
    The data suggest that the topological susceptibility near the chiral limit
    is suppressed to the level of $<10$ MeV as $\chi_t^{1/4}/T\sim m$.
  }
  \label{fig:chitT4throot}
\end{figure}

In Fig.~\ref{fig:chitT}, we present the
data at various temperatures.
The volume size is fixed to $L=32$ (2.4 fm), except for
data at $T=195$ MeV ($\beta=4.24$) (2.7 fm) denoted by diamonds
and those at $T=330$ MeV (3.6 fm) by crosses.
Here the filled symbols are those obtained with reweighting
from the ensemble at a higher mass point shown by open symbols.
Even on the configurations where topology fluctuates frequently,
the reweighted results decrease towards the chiral limit,
which is consistent with the non-reweighted data.
See Tab.~\ref{tab:prms} for the ensembles
and masses to which we applied the mass reweighting.


In order to focus on the region near the chiral limit and
compare the scale of $\chi_t$ compared to the temperature,
we plot the same data in Fig.~\ref{fig:chitT4throot} 
taking the fourth root of $\chi_t$ and normalizing it by $T$.
It suggests that the topological susceptibility near the chiral limit
is suppressed to the level of $O(10)$ MeV with a power $\sim m^4$ (or $\chi_t^{1/4}\sim m$).
The results are not precise enough
to determine if $\chi_t$ goes to zero at finite quark mass,
as predicted in \cite{Aoki:2012yj}.


\clearpage
\subsection{Axial $U(1)$ susceptibility}

Let us investigate a more direct measure of
the violation of the axial $U(1)$ symmetry, {\it i.e.}
the axial $U(1)$ susceptibility, defined by
the difference between the pseudoscalar ($\pi$) and scalar ($\delta$)
correlators integrated over spacetime,
\begin{eqnarray}
\Delta(m) = \sum_x \left[\langle \pi(x)\pi(0)\rangle-\langle \delta(x)\delta(0)\rangle\right],
\end{eqnarray}
where the ensemble average is taken at a finite quark mass $m$.
For the overlap-Dirac operator, we can express $\Delta(m)$
using the spectral decomposition (see \cite{Cossu:2015kfa} for the details):
\begin{eqnarray}
  \label{eq:Deltaeigen}
\Delta(m)=  \frac{1}{V(1-m^2)^2}\left\langle \sum_{\lambda_m} \frac{2m^2(1-\lambda_m^2)^2}{\lambda_m^4}\right\rangle,
\end{eqnarray}
where $\lambda_m$'s are the eigenvalues of $H_{ov}(m)\equiv \gamma_5D_{ov}(m)$.
Although this equality holds even in a finite volume, we must take
the thermodynamical limit before taking the chiral limit.

\begin{figure}[tbp]
  \centering
  \includegraphics[width=14cm]{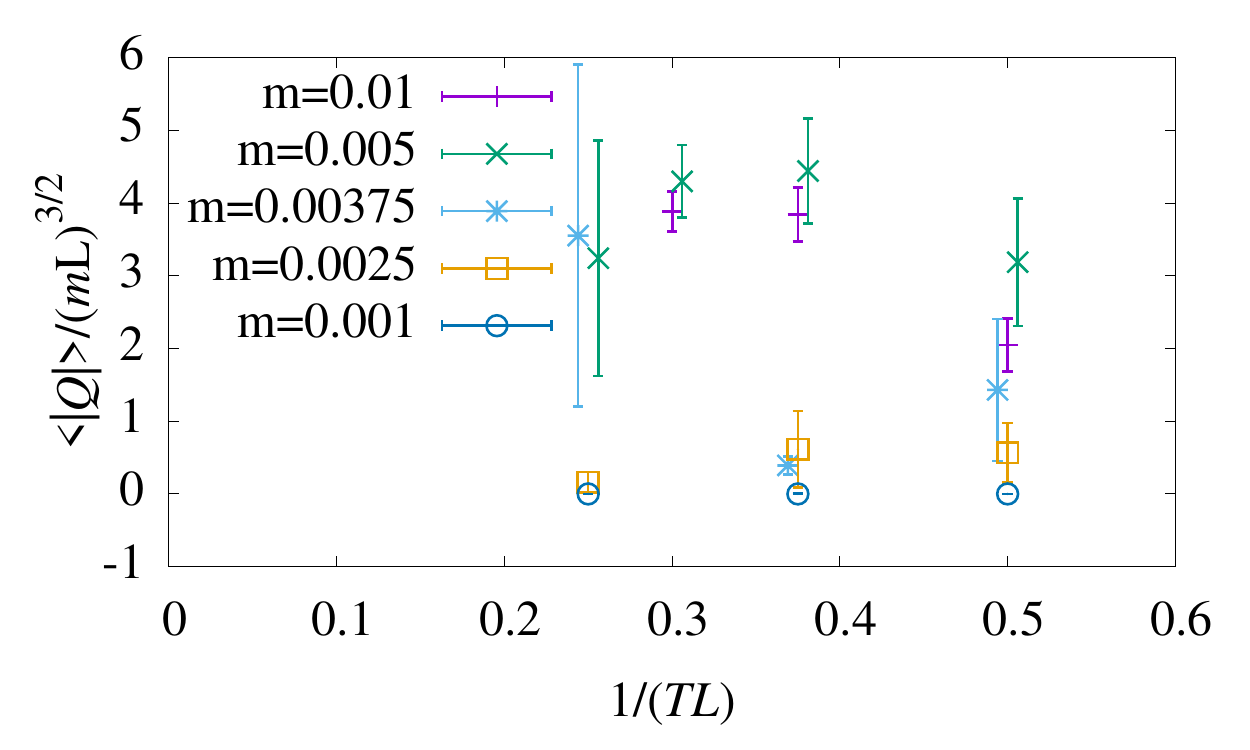}
  \caption{
    Finite size scaling of the chiral zero modes' effect.
    Data at $T=220$ MeV are shown.
    The expected $L^{3/2}$ scaling of $\langle |Q|\rangle$ is saturated for $1/LT<0.4$.
  }
  \label{fig:N0}
\end{figure}

In our previous study \cite{Tomiya:2016jwr}, we found that
the contribution from chiral zero modes is quite noisy.
As an alternative, we subtract the zero-mode contribution, for which $\lambda_m=m$, and define
\begin{equation}
  \label{eq:Deltabareigen}
\bar{\Delta}(m) \equiv \Delta(m)-\frac{2\langle |Q|\rangle}{m^2(1-m^2)^2V},
\end{equation}
where $Q$ is the index of the overlap-Dirac operator.
We remind the reader that the index $Q$ is equal to the
topological charge of the gauge field.
This subtraction is justified because in the thermodynamical limit,
while at a fixed temperature,
$\langle |Q|\rangle$ scales as $V^{1/2}$ ($=L^{3/2}$), and thus the zero-mode contributions
vanish in the large volume limit as $1/V^{1/2}$.
We numerically confirm this scaling at $T=220$ MeV as presented in Fig.~\ref{fig:N0}.
The  $L^{3/2}$ scaling of $\langle |Q|\rangle$ looks saturated for $1/(TL)<0.4$.
Therefore, $\bar{\Delta}(m)$ in the thermodynamical limit
coincides with $\Delta(m)$.

In this work, we further refine the observable by removing the UV divergence.
From a simple dimensional analysis of the spectral expression
in Eq.~(\ref{eq:Deltaeigen}), the valence quark mass $m_v$ dependence of
$\bar{\Delta}(m)$ can be expanded as
\begin{eqnarray}
  \frac{A}{m_v^2} + B + m_v^2 C + O(m_v^4),
\end{eqnarray}
and $C$ has a logarithmic UV divergence. What we are interested in
is the divergence free piece $\frac{A}{m_v^2} + B$ and
if it is zero or not in the chiral limit.
Note that $A$ and $B$ contain a sea quark mass $m_{sea}$ dependence,
and $A$, in particular, should be suppressed as $m_{sea}^2$,
at least, to avoid possible IR divergence in the limit of $m_{sea}=m_v\to 0$.
Measuring $\bar{\Delta}(m)$ at three different valence masses
$m_{1,2,3}$, we extract the UV finite quantity:
\begin{eqnarray}
  \label{eq:DeltaUVsubt}
  \bar{\Delta}^{UV subt.} &=&
  \frac{m_2^2 m_3^2}{m_2^2-m_3^2}
  \left[\frac{\bar{\Delta}(m_1)-\bar{\Delta}(m_2)}{m_1^2-m_2^2}-\frac{\bar{\Delta}(m_1)-\bar{\Delta}(m_3)}{m_1^2-m_3^2}\right]
  \nonumber\\&&+\frac{(m_1^2+m_2^2)(m_1^2+m_3^2)}{m_3^2-m_2^2}
  \left[\frac{m_1^2\bar{\Delta}(m_1)-m^2_2\bar{\Delta}(m_2)}{m_1^4-m_2^4}-\frac{m_1^2\bar{\Delta}(m_1)-m^2_3\bar{\Delta}(m_3)}{m_1^4-m_3^4}\right],
\end{eqnarray}
while fixing the sea quark mass $m$. By choosing $m_1=m$ and $m_{2,3}$ in its vicinity,
one can easily confirm that $\bar{\Delta}^{UV subt.}(m)=A/m^2+B+O(m^4)$.
In this work, we choose $m_2=0.95 m$ and $m_3=1.05 m$.

\begin{figure}[tbp]
  \centering
  \includegraphics[width=8cm]{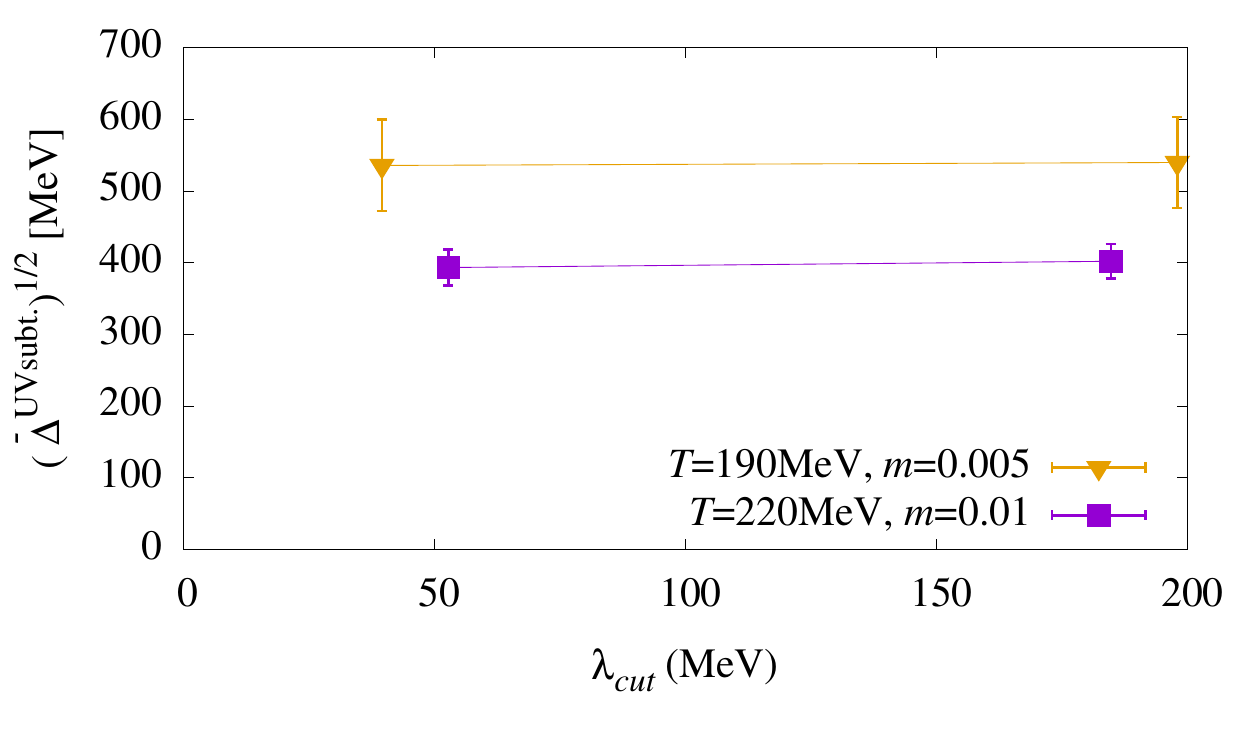}
  \includegraphics[width=8cm]{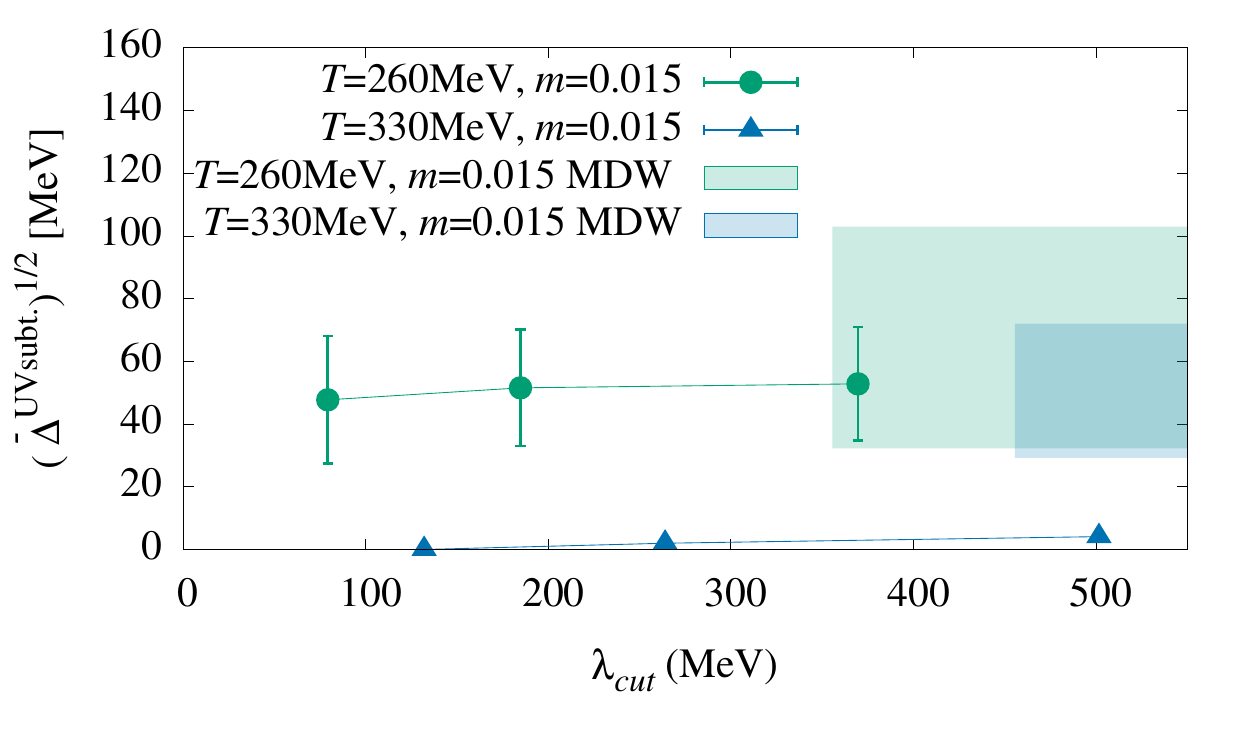}
  \caption{
    Cut-off $\lambda_{cut}$ dependence of $[\bar{\Delta}^{UV subt.}]^{1/2}$.
    The left panel shows data at $T=190$ MeV and $220$ MeV
    and the right is for $T=260$ and 330 MeV.
    The lower three temperatures show a good saturation
    but the data at $T=330$ MeV is monotonically increasing and
    undershoot the band, which represents the stochastic
    estimates using the
     M\"obius domain-wall Dirac operator.
  }
  \label{fig:lowmode-Delta}
\end{figure}

We compute $\bar{\Delta}(m)$ through the expressions in Eqs.~(\ref{eq:Deltaeigen}) and (\ref{eq:Deltabareigen})
truncating the sum at a certain upper limit $\lambda_{cut}$ (around $180-500$ MeV).
We then use Eq.~(\ref{eq:DeltaUVsubt}) to obtain the UV subtracted susceptibility.
Figure~\ref{fig:lowmode-Delta} shows the $\lambda_{cut}$ dependence of $[\bar{\Delta}^{UV subt.}]^{1/2}$,
where the left panel shows data at $T=190$ MeV and $220$ MeV
and the right is for $T=260$ and 330 MeV.
The data at lower three temperatures are well saturated at $\lambda_{cut}\sim$ 50 MeV,
while the data at $T=330$ MeV show a monotonic increase though its magnitude is small.
The shadowed bands are stochastic estimates
of the two-point functions using the  M\"obius domain-wall Dirac operator with
three different valence quark masses.
This estimates contain contributions from all possible modes under the lattice cut-off,
and the consistency between
the two methods at $T=260$ MeV supports our observation
that the low-mode approximation is good for $T=260$ MeV and below.
The data also show the consistency between the overlap and
M\"obius domain-wall fermion formulations at this temperature,
in contrast to disagreement at lower temperatures (see below).
As shown in the Dirac eigenvalue density,
the eigenvalues are pushed up for higher temperatures,
which makes the low-mode approximation worse,
but makes the violation of the Ginsparg-Wilson relation less crucial.
In the following analysis, we use the 
stochastic  M\"obius domain-wall results for $T=330$ MeV
and the low-mode approximation of the overlap fermion
for the other lower temperatures.

\begin{figure}[tbp]
  \centering
  \includegraphics[width=14cm]{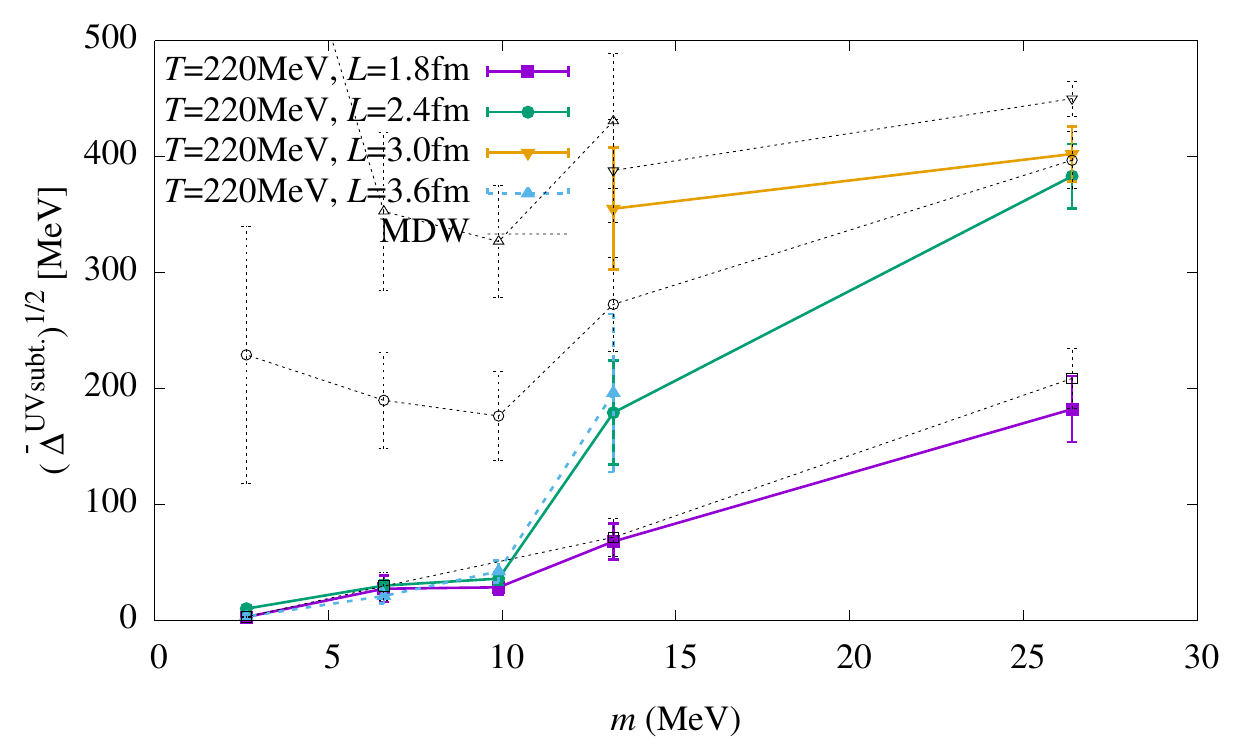}
  \caption{
    $[\bar{\Delta}^{UV subt.}]^{1/2}$ as a function of $m$.
    Data at $T=220$ MeV with four different volumes are shown.
    at $T=220$ MeV are shown.
    Filled symbols with solid lines are data of reweighted overlap fermion,
    those with dashed lines are at $L=48$ whose statistics may be not good enough
    after the reweighting,
    and dotted open symbols are those of the M\"obius domain-wall fermion.
  }
  \label{fig:Delta220}
\end{figure}

In Fig.~\ref{fig:Delta220}, the results for $[\bar{\Delta}^{UV subt.}]^{1/2}$
at $T=220$ MeV are shown.
Filled symbols with solid lines are data of reweighted overlap fermion
and dashed open symbols are those of the M\"obius domain-wall fermion.
As reported in \cite{Tomiya:2016jwr}, the M\"obius domain-wall fermion results
deviate from due to the sensitivity of the observable
to the violation of the Ginsparg-Wilson relation.
Except for the heaviest two points with $L=1.8$ fm
where the aspect ratio $LT$ is 2,
the data are consistent among four different volumes.
The axial $U(1)$ anomaly is strongly suppressed
to a few MeV level at the lightest quark mass.

\begin{figure}[tbp]
  \centering
  \includegraphics[width=14cm]{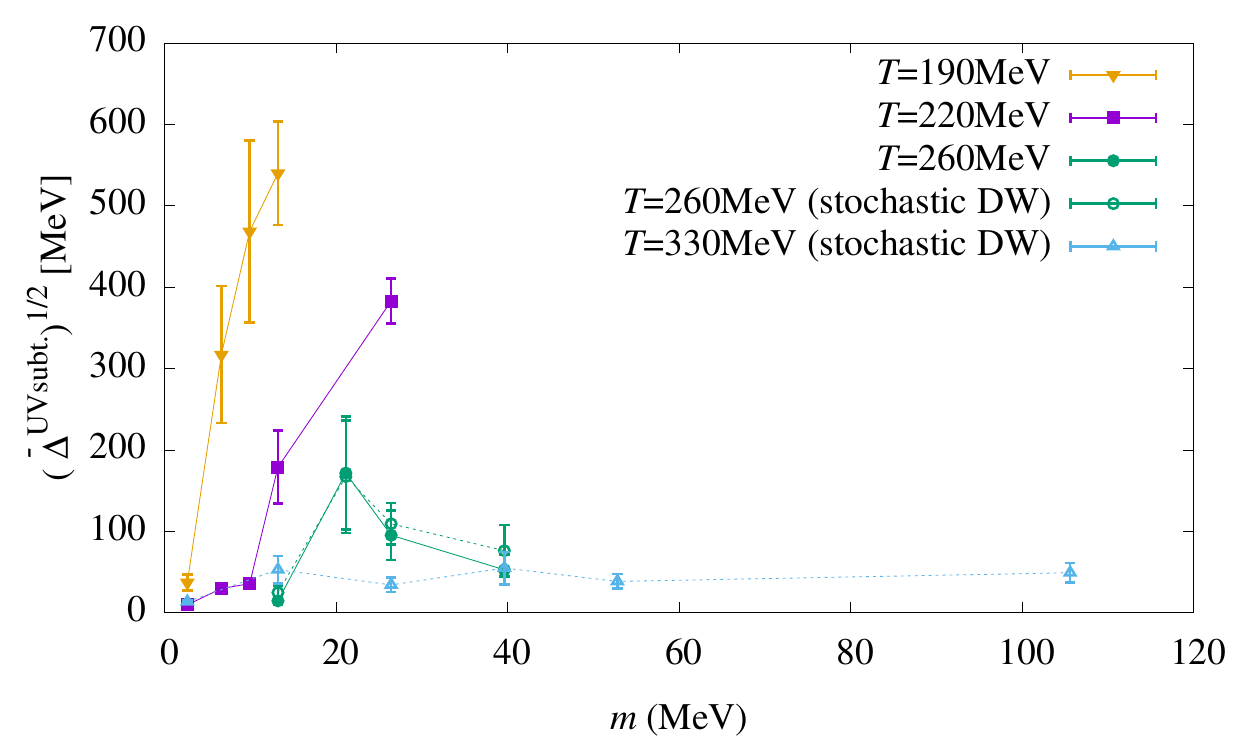}
  \caption{
    The same as Fig.~\ref{fig:Delta220} but at different temperatures.
  }
  \label{fig:DeltaT}
\end{figure}

This strong suppression is also seen at
different temperatures, as presented in Fig.~\ref{fig:DeltaT}.
For $T=260$ MeV, the overlap and
M\"obius domain-wall fermion results agree well, in contrast to the lower temperatures.
The chiral limit of $[\bar{\Delta}^{UV subt.}]^{1/2}$ looks consistent with zero,
and the value near the physical point is $20$ MeV, at most.

In Tab.~\ref{tab:results}, we summarize the results for
the lowest bin of the eigenvalue density or $\rho(\lambda=0)$,
    $\langle Q^2\rangle=\chi_t V$, and axial $U(1)$ susceptibility $\bar{\Delta}^{UV subt.}$.


\renewcommand{\arraystretch}{0.5}
\begin{table}[tbp]
  \centering
  \begin{tabular}{ccccccccc}
    \hline\hline
    $\beta$ & $L^3\times L_t$ & $T$(MeV) & $m$ & $\rho(\lambda=0)$ & $\langle Q^2\rangle$ & $\bar{\Delta}^{UV subt.}$ \\
    \hline
    4.24 & $32^3\times 12$ & 195 & 0.0025 & 0.00211(91)& 0.084(37)
    \\
        &  &  & 0.005 & 0.0134(22)& 0.575(83)
    \\
        &  &  & 0.01 & 0.0393(69)& 3.1(10)
    \\
 \hline
 4.30 & $32^3\times 14$ & 190 & 0.001& 0.9(9)$\times 10^{-9}$& 4(4)$\times 10^{-8}$& 0.00020(11)
 \\
   &  &  & 0.0025& 0.0037(11)& 0.113(29)& 0.0144(77)
 \\
   &  &  & 0.00375& 0.0126(25)& 0.69(17)& 0.031(15)

 \\
 &  &  & 0.005& 0.0113(19)& 0.442(83)& 0.0419(98)
 \\
\cline{2-7}
 & $32^3\times 12$ & 220 & 0.001& 0.8(8)$\times 10^{-7}$& 4(4)$\times 10^{-6}$& 1.4(9)$\times 10^{-5}$
\\
  &  &  & 0.0025& 0.00035(22)& 0.014(12)& 0.000128(35)
\\
  &  &  & 0.00375& 0.000262(88)& 0.0110(36)& 0.000185(57)
\\
  &  &  & 0.005& 0.0064(14)& 0.367(74)& 0.0046(23)
\\
  &  &  & 0.01& 0.0160(22)& 1.13(16)& 0.0211(31)
\\
\cline{2-7}
&   $32^3\times 10$ & 260 & 0.005& 0.00104(76)& 0.043(20)& 3(2)$\times 10^{-5}$
\\
  &  &  & 0.008& 0.0031(11)& 0.122(38)& 0.0042(34)
\\
&  &  & 0.01& 0.0047(13)& 0.232(71)& 0.00130(83)
\\
      &  &  & 0.015& 0.0057(22)& 0.251(96)& 0.00040(27)
\\
 \cline{2-7}
 &   $32^3\times 8$ & 330 & 0.001 & 0(0)& 0(0)& 3(6)$\times 10^{-5}$
 \\
 &  &  & 0.005 & 1.2(9)$\times 10^{-5}$& 0.00049(39)& 0.00040(26)
 \\
       &  &  & 0.01& 6(2)$\times 10^{-5}$& 0.0024(14)& 0.000169(87)
 \\
       &  &  & 0.015& 0.00074(62)& 0.044(28)& 0.00043(31)
 \\
        &  &  & 0.02& 0.0025(10)& 0.099(41)& 0.000215(99)
 \\
       &  &  & 0.04& 0.0044(16)& 0.508(97)& 0.00035(17)
 \\
\hline\hline
  \end{tabular}
  \caption{The results for the lowest bin of the eigenvalue density or $\rho(\lambda=0)$,
    $\langle Q^2\rangle=\chi_t V$, and axial $U(1)$ susceptibility $\bar{\Delta}^{UV subt.}$.
  For $T=330$ MeV, the stochastic  M\"obius domain-wall results are listed. }
  \label{tab:results}
\end{table}

\clearpage
\subsection{Meson correlators}

In the previous subsection we
studied the difference between
the pseudoscalar and scalar two-point correlation
function integrated over the whole lattice.
Since we have subtracted the short-range UV divergent
part, the quantity is essentially probing
physics at the scale of our lattice size $L$,
which is sensitive to the near zero modes.
In this subsection, we investigate
the mesonic two point correlation function itself,
which must contain shorter-range information of QCD,
as our fitting range is typically $L/4$.

We measure the spatial correlator in the $z$ direction
\begin{eqnarray}
  C_\Gamma(z) = - \sum_{x,y,t} \langle \mathcal{O}_\Gamma(x,y,z,t) \bar{\mathcal{O}}_\Gamma (0,0,0,0)\rangle, 
\end{eqnarray}
with $\mathcal{O}_\Gamma = \bar{q}\vec\tau\Gamma q$.
Here $\vec\tau$ are the generators in the flavor space.
For $\Gamma$ we choose $\gamma_5$($PS$), $1$($S$), $\gamma_{1,2}$($V$),
$\gamma_5\gamma_{1,2}$($A$),
$\gamma_4\gamma_3$($T_t$) and $\gamma_5\gamma_4\gamma_3$($X_t$).
In this work, we focus on $T_t$, and $X_t$ channels,
which are related by the axial $U(1)$ transformation,
as well as the $V$ and $A$ channels to check the recovery of the $SU(2)_L\times SU(2)_R$ symmetry.
We find that the $S$ correlator is too noisy
to extract the ``mass'' and compare
it with that of $PS$.
For other channels, we will report elsewhere.

Since this quantity represents shorter distance physics than the axial $U(1)$ susceptibility obtained
from integration over whole lattice,
the violation of the Ginsparg-Wilson relation enhanced by near-zero modes found in \cite{Tomiya:2016jwr} is less severe.
Therefore, we employ the  M\"obius domain-wall fermion formalism without reweighting.
In order to improve the statistics, rotationally equivalent
directions are averaged.
Also, low-mode averaging~\cite{Giusti:2004yp,DeGrand:2004qw}
using 40 lowest eigenmodes of $H_{DW}(m)$
is performed with four equally spaced source points at temperature $T=220$ MeV.


For the channels other than $S$, the signal is good enough
to extract the asymptotic behavior of $C_\Gamma(z)$ at large $z$,
which should contain the information of the screening mass.
However, the correlator at high temperatures may not behave
like a single exponential even at large $z$.
To circumvent this, recent studies often apply
multi-state fits and introduce various
source types to extract ground-state values,
{\it e.g.} in~\cite{Bazavov:2019www}.

The asymptotic behavior depends on the structure of
the spectral functions (in the spatial direction) for each channel $\Gamma$:
\begin{eqnarray}
  C_\Gamma(z) = \int d\omega \rho_\Gamma(\omega) \int \frac{dp_z}{2\pi} \frac{2\omega e^{ip_z z}}{p_z^2+\omega^2}
  = \int d\omega \rho_\Gamma(\omega)e^{-\omega z}.
\end{eqnarray}
When the spectral function $\rho_\Gamma(\omega)$ starts from a
series of delta functions, which represents isolate poles, 
$C_\Gamma(z)$ at large $z$ is dominated by a single exponential.
On the other hand, if the correlator is described by deconfined
two quarks, $\rho_\Gamma(\omega)$ is a continuous function of $\omega$,
provided that the volume is sufficiently large.
Let us assume that $\rho_\Gamma(\omega)$ becomes nonzero at
a threshold $2\bar{m}$, where $\bar{m}$ is
a constituent screening quark mass.
For large $z$, we can expand $\rho_\Gamma(\omega)$ as $\theta(\omega-2\bar{m})(c_0 +c_1 (\omega-2\bar{m}) +\cdots)$ (with a step function $\theta$),
which results in $C_\Gamma(z)\sim \exp(-2\bar{m}z)(c_0/z+O(1/z^2))$.
In the Appendix~\ref{app:spec}, we show that this form of the spectral function
is indeed realized in the free two quark propagators.
The essential difference from the single exponential is,
thus, the factor $1/z$.
In this study we therefore apply two types of fitting functions:
the standard cosh function,
\begin{eqnarray}
  \label{eq:cosh}
A_\Gamma \frac{\cosh(m_\Gamma(z-L/2))}{\sinh(m_\Gamma L/2)},
\end{eqnarray}
and the  two-quark-inspired function (2q),
\begin{eqnarray}
  \label{eq:2qfunc}
  B_\Gamma\left(\frac{e^{-m'_\Gamma z}}{m'_\Gamma z}+\frac{e^{-m'_\Gamma(L-z)}}{m'_\Gamma(L-z)}\right).
\end{eqnarray}
In the free quark limit, we obtained more complete form
of the two-quark propagations for each channel \cite{Rohrhofer:2017grg}.
Note that in this limit, $\bar{m}=\pi T$, which is the lowest Matsubara mass.
It is, therefore, interesting to see how much $m'_\Gamma$ is close to $2\pi T$
at our simulated temperature.

\begin{figure*}[tbh]
  \centering
  \includegraphics[width=8cm]{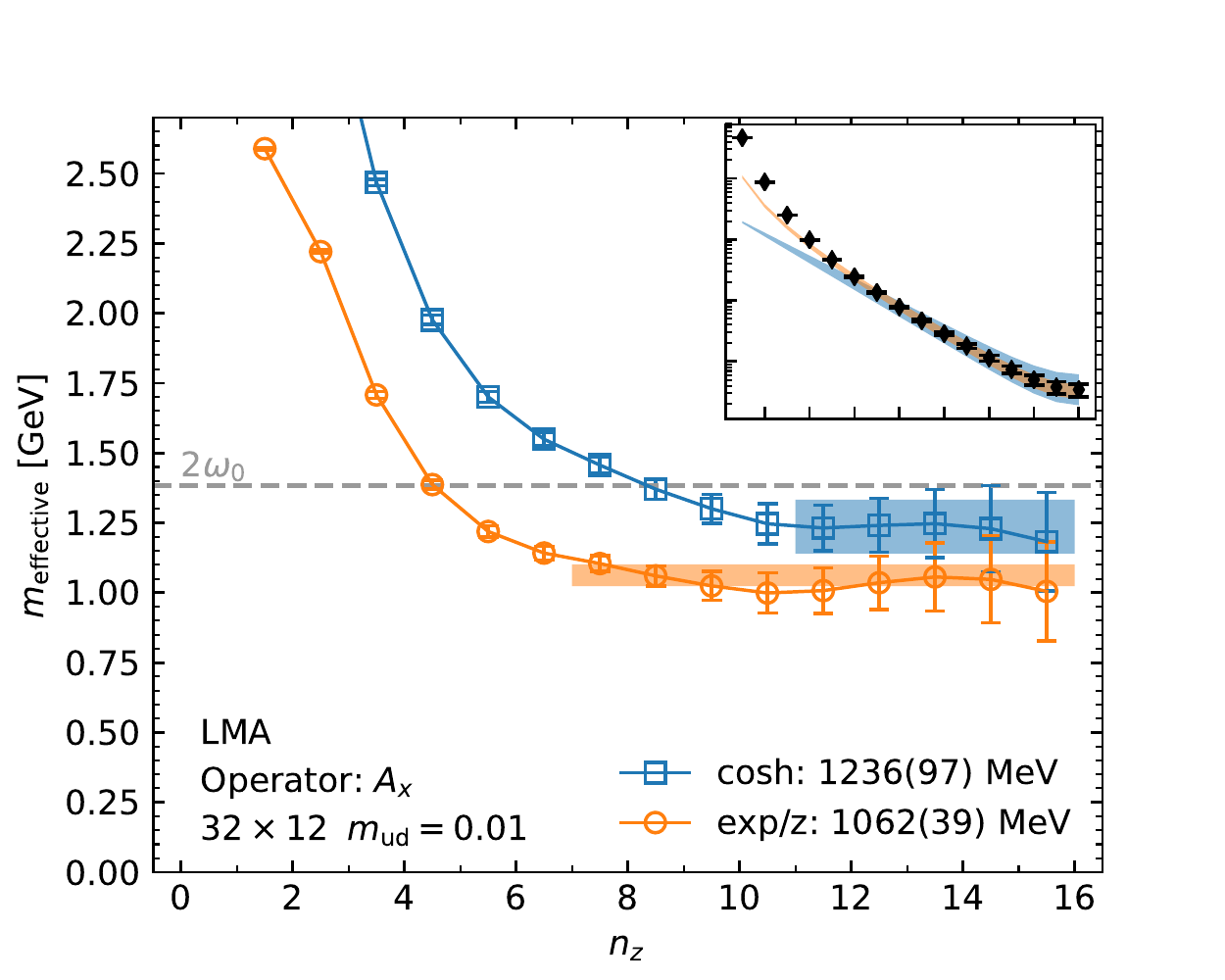}
  \includegraphics[width=8cm]{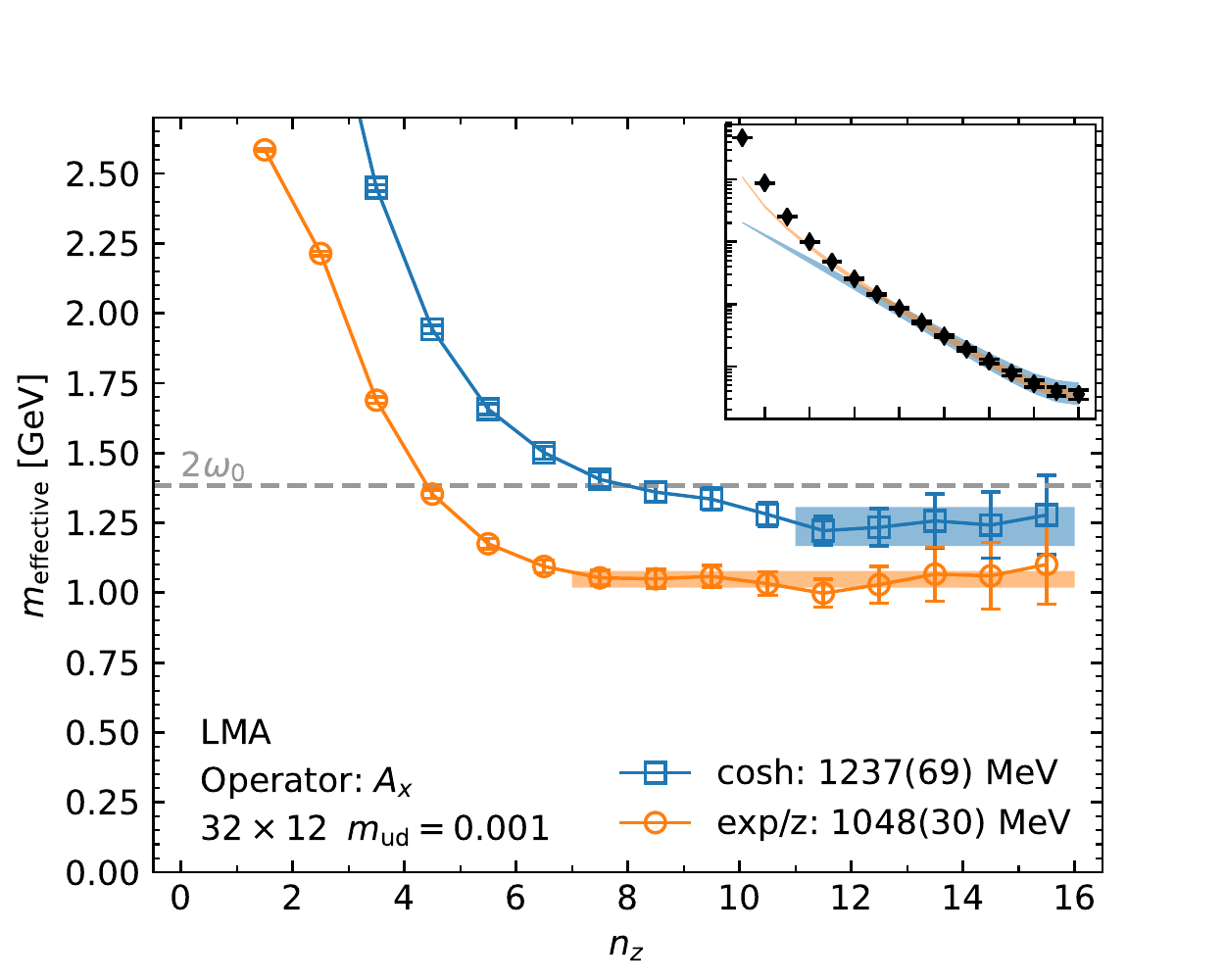}
  \includegraphics[width=8cm]{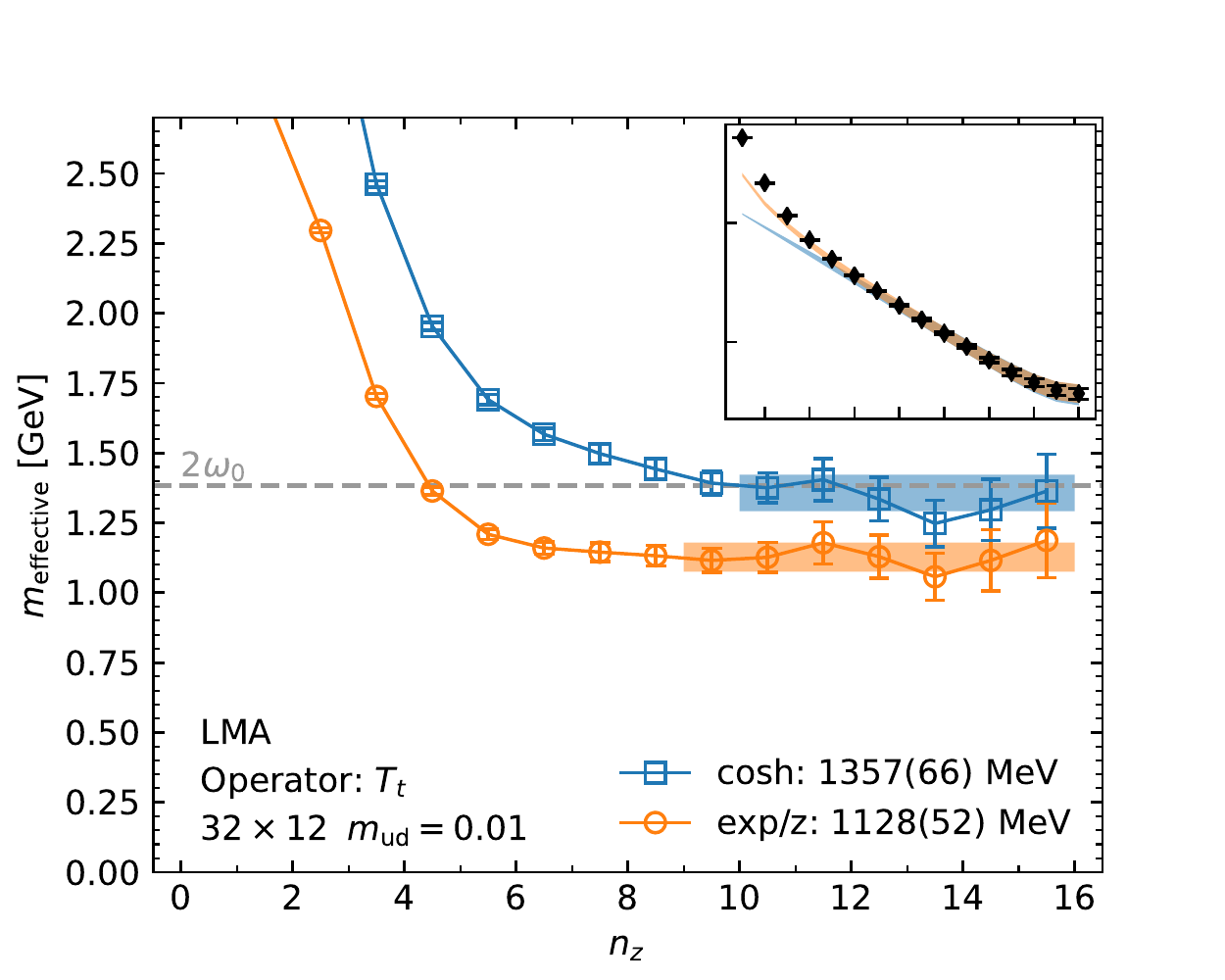}
   \includegraphics[width=8cm]{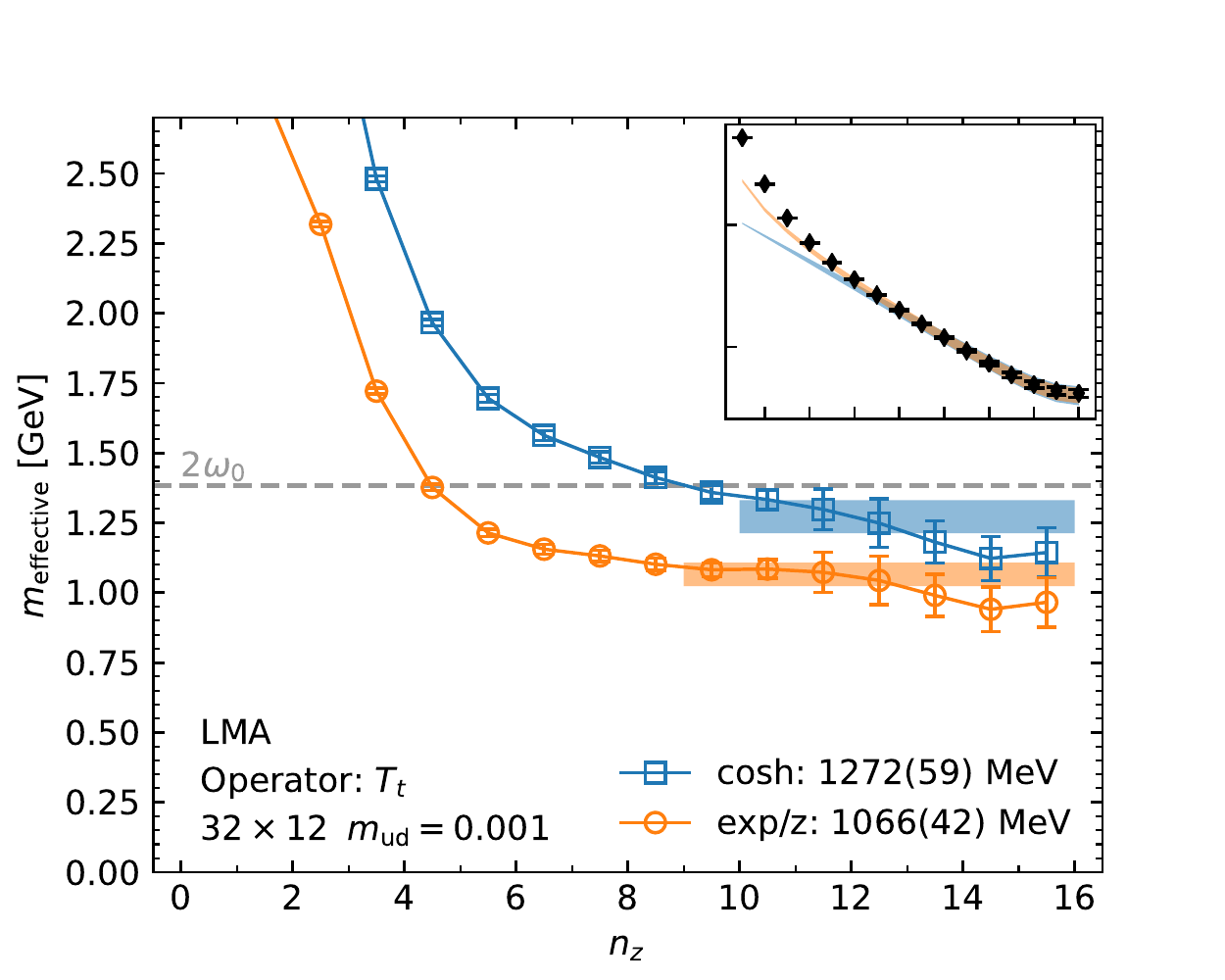}
  \caption{
    Cosh (squares)  and ``2-quark''(triangles) effective masses at $T=220$ MeV.
    The top panels show the data from $A$ correlators, while the bottom panels
    are $T_t$ correlators.
    The left panels are at the heaviest simulated mass $m=0.01$ and
    the right ones are at the lightest $m=0.001$.
    The band indicates the fit result and its range.
    The dashed line shows the Matsubara frequency $\pi T$ times 2 expected for two free quarks.
  }
  \label{fig:Meffectivemass}
\end{figure*}

In order to compare the above fitting functions, it is helpful to plot
their effective masses
$m_\Gamma(z)$ and $m'_\Gamma(z)$ defined as the solutions of
\begin{eqnarray}
  \frac{\cosh(m_\Gamma(z)(z-L/2))}{\cosh(m_\Gamma(z)(z+1-L/2))}
  &=& \frac{C_\Gamma(z)}{C_\Gamma(z+1)},
\end{eqnarray}
and
\begin{eqnarray}
  \frac{\frac{e^{-m'_\Gamma(z) z}}{m'_\Gamma(z) z}+\frac{e^{-m'_\Gamma(z)(L-z)}}{m'_\Gamma(z)(L-z)}}{\frac{e^{-m'_\Gamma(z) (z+1)}}{m'_\Gamma(z) (z+1)}+\frac{e^{-m'_\Gamma(z)(L-z-1)}}{m'_\Gamma(z)(L-z-1)}}
  &=& \frac{C_\Gamma(z)}{C_\Gamma(z+1)},
\end{eqnarray}
respectively, with the lattice data of $C_\Gamma(z)/C_\Gamma(z+1)$.
If the fitting form is
good, the effective mass converges to a constant
at a shorter value of $z$.
In Fig.~\ref{fig:Meffectivemass}, we plot
typical effective mass plots at $T=220$ MeV.
Square symbols are data for $m_\Gamma(z)$ and
the circles are those for $m'_\Gamma(z)$.
The top two panels show the data obtained from $A$ correlators,
while the bottom panels are from $T_t$ correlators.
The left panels are at the heaviest simulated mass $m=0.01$ and
the right ones are at the lightest $m=0.001$.
The band indicates the fitting result and its range.
We find that the 2q function shows longer and more stable plateux.

It is interesting to see that the plateaux are located at
a mass lower than $2\pi T$
which is indicated by dashed lines.
While we plan to give a detailed analysis in another publication~\cite{JLQCD:2020},
we show the screening mass $m'$ obtained by a fit to
the 2q formula Eq.~(\ref{eq:2qfunc}).
The results are listed in Tab.~\ref{tab:screeningmasses_mesons}.

\begin{figure*}[tbh]
  \centering
  \includegraphics[width=14cm]{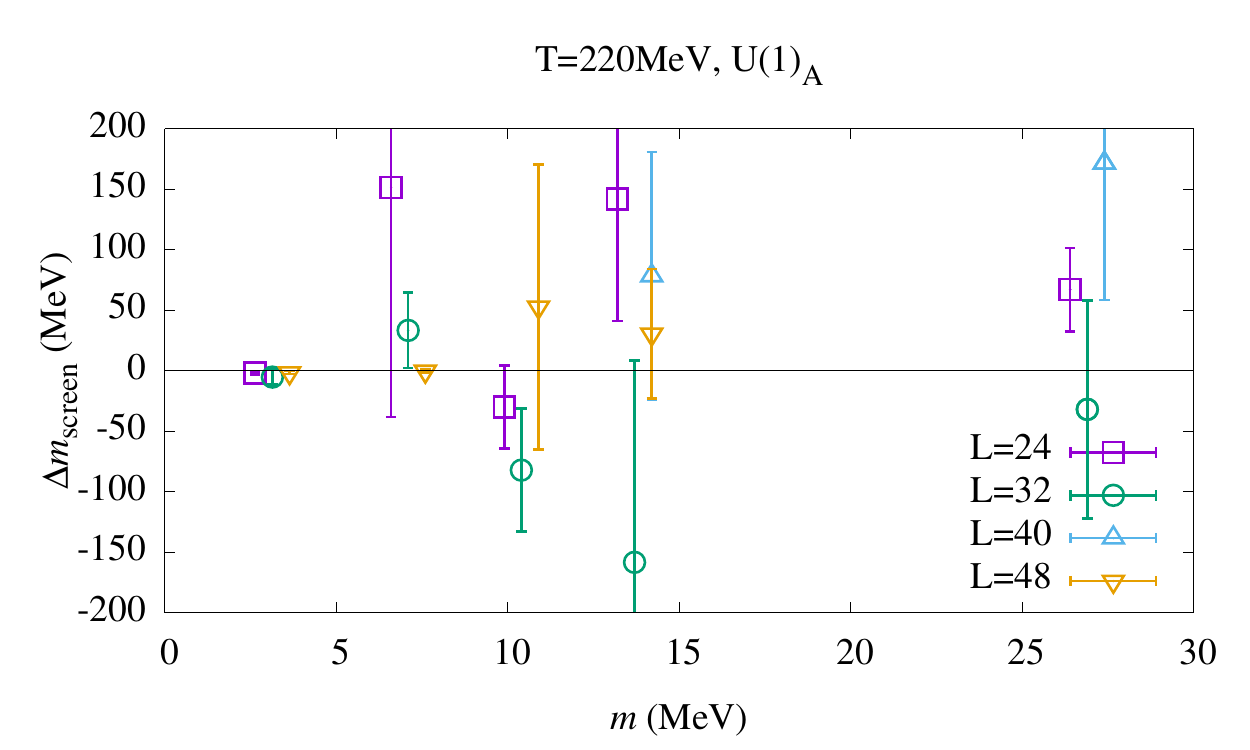}
    \caption{
      Difference of the fitted $m'$ between $T_t$ and $X_t$ correlators,
      which are connected by the $U(1)_A$ rotation.
      The data at $T=220$ MeV at different volumes are shown.
    }
    \label{fig:MesondifU1}
  \includegraphics[width=14cm]{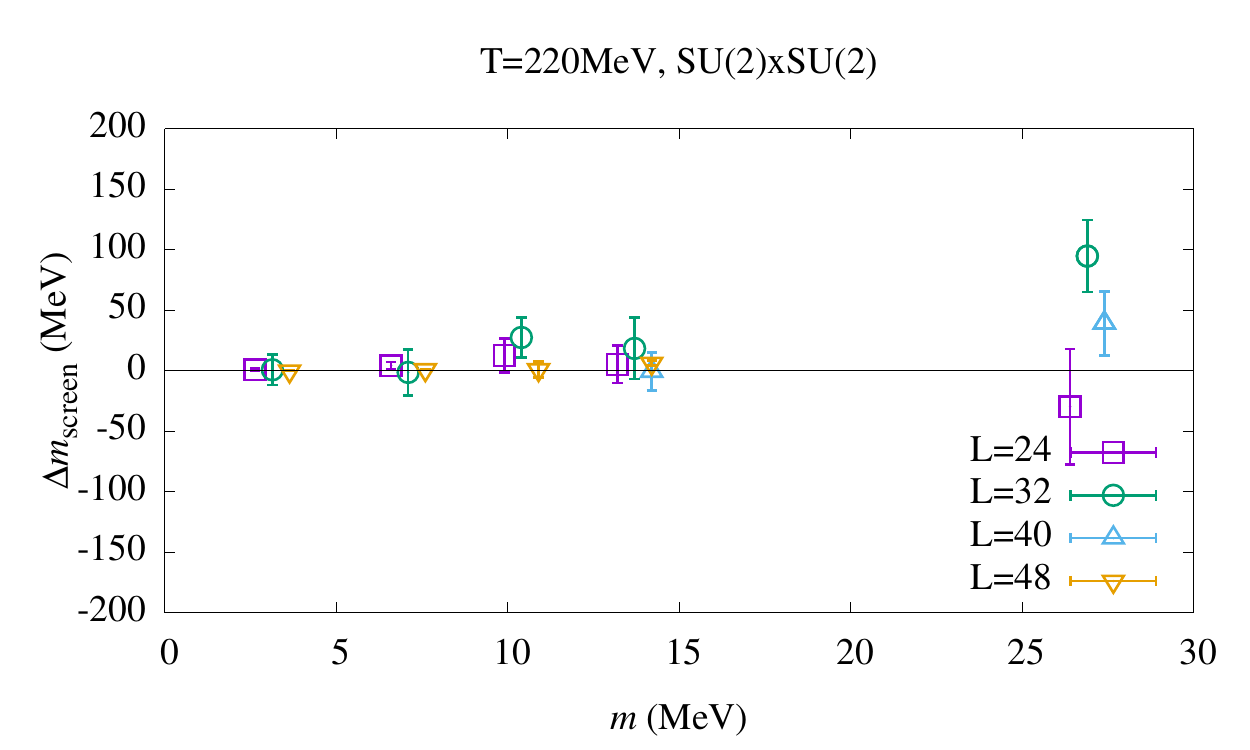}
  \caption{
    The same as Fig.~\ref{fig:MesondifU1} but between $A$ and $V$ correlators,
    which shows the $SU(2)_L\times SU(2)_R$ symmetry.
  }
  \label{fig:MesondifSU2}
\end{figure*}

In Fig.~\ref{fig:MesondifU1}, we plot the difference of the fitted $m'$ between
$T_t$ and $X_t$ correlators at $T=220$ MeV at different volumes.
They are connected by the $U(1)_A$ rotation, and therefore, the
difference, denoted by $\Delta m_{\rm screen}$, is a probe of the axial $U(1)$ symmetry.
For the reference, we also plot in Fig.~\ref{fig:MesondifSU2} the results for
the difference between $A$ and $V$ channels, which is an indicator for the
$SU(2)_L\times SU(2)_R$ symmetry.

\begin{figure*}[tbh]
  \centering
  \includegraphics[width=8cm]{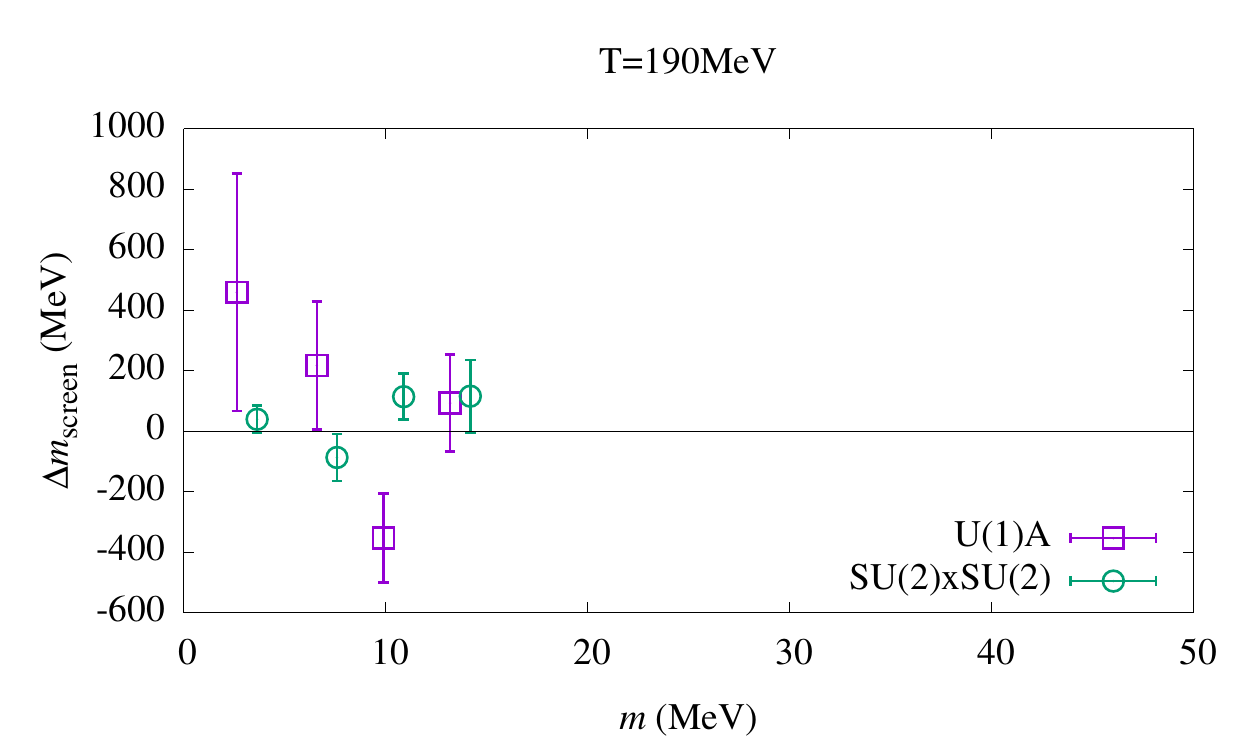}
  \includegraphics[width=8cm]{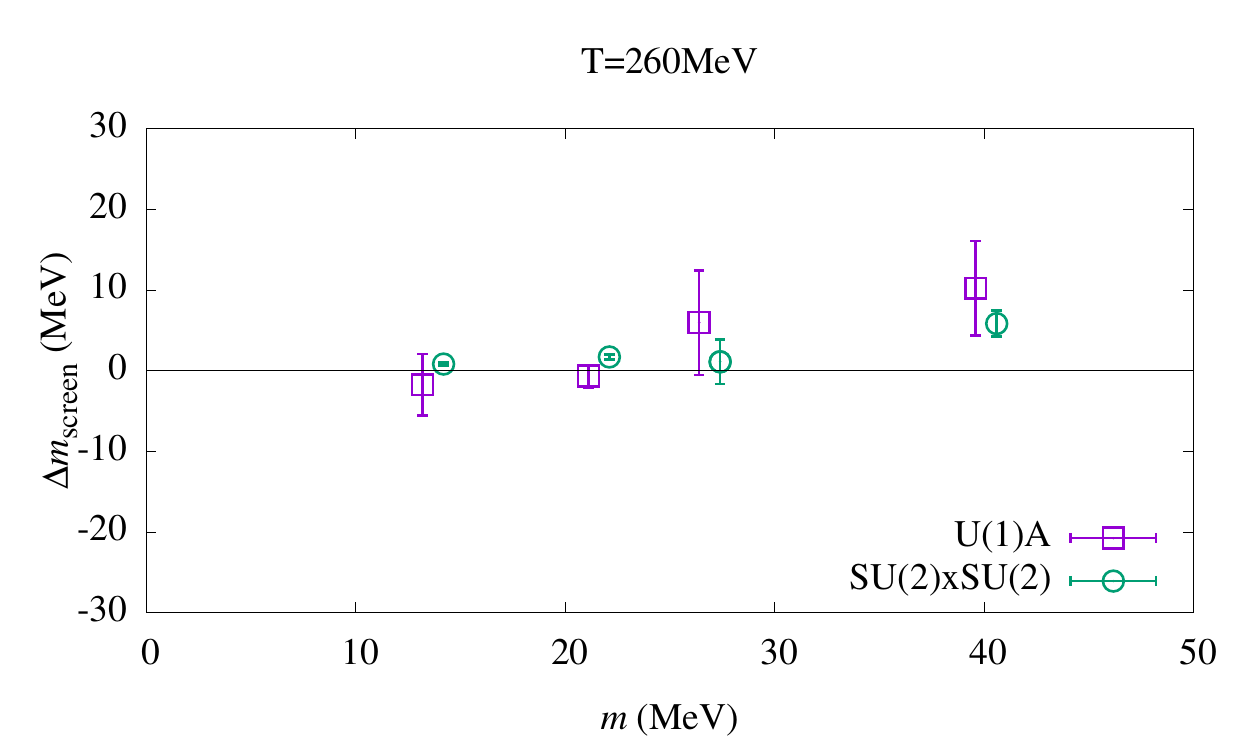}
  \includegraphics[width=8cm]{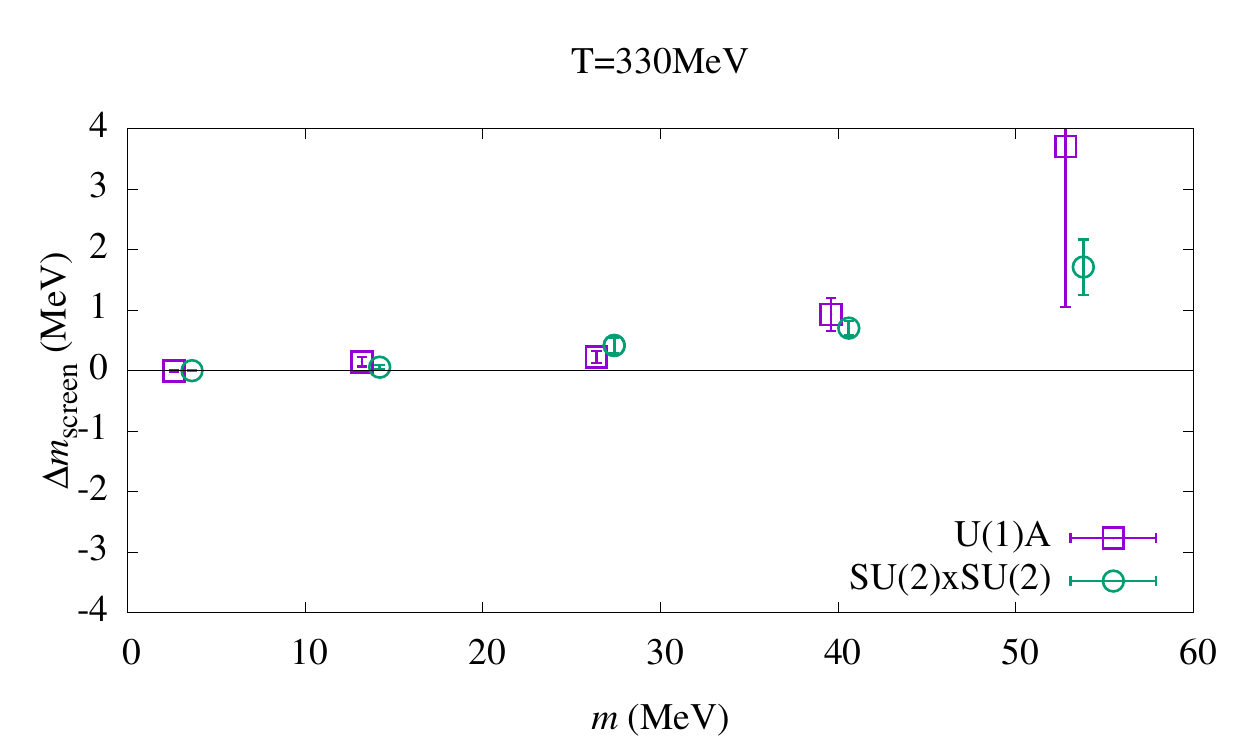}
  \caption{
    Difference of the fitted $m'$ between $T_t$ and $X_t$ correlators (squares),
    and that between $A$ and $V$ correlators (circles).
  }
  \label{fig:MesondifT}
\end{figure*}

Although the $U(1)$ data at heavier quark masses are noisier than those
for $SU(2)_L\times SU(2)_R$, their chiral limit looks consistent with zero,
and the central values are only a few MeV, at the lightest quark mass.
We note that their individual mass is $\sim 1$ GeV.
Therefore, the axial $U(1)$ symmetry relation is satisfied at a sub-\% level.
This behavior is also seen at different temperatures, as shown in Fig.~\ref{fig:MesondifT},
except for $T=190$ MeV (but they are still consistent with zero with large error bars).
The disappearance of the axial $U(1)$ anomaly is
consistent with other observables obtained using the reweighted overlap fermions.

\renewcommand{\arraystretch}{0.5}
\begin{table}[tbp]
  \centering
  \begin{tabular}{cccc|cccccc}
\hline\hline
    &  &  &  & \multicolumn{6}{c}{$m'$(MeV)}\\
    $\beta$ & size & $T$(MeV) & $ma$ & $PS$ & $S$ & $V$ & $A$ & $T_t$ & $X_t$  \\
\hline
 4.30 
 & $32^3 \times 14$ & 190 & 0.001 & & 172(104)  & 832(66)  & 872(62)  & 749(171)  & 1209(249)  \\
 &  &  & 0.0025 & 44(41)  & & 1015(216)  & 928(214)  & 802(100)  & 1020(129)  \\
 &  &  & 0.00375 & 118(20)  & & 747(61)  & 862(86)  & 1111(61)  & 759(125)  \\
 &  &  & 0.005 & 147(24)  & & 806(99)  & 923(151)  & 942(82)  & 1036(123)  \\
\cline{2-10}
 & $24^3 \times 12$ & 220 & 0.001 & 482(31)  & 482(32)  & 1027(41)  & 1028(41)  & 1088(22)  & 1086(22)  \\
 &  &  & 0.0025 & 413(47)  & 510(64)  & 1074(91)  & 1078(92)  & 1015(94)  & 1166(108)  \\
 &  &  & 0.00375 & 512(31)  & 590(59)  & 1055(44)  & 1068(46)  & 1138(35)  & 1108(31)  \\
 &  &  & 0.005 & 386(39)  & 640(237)  & 1103(54)  & 1109(56)  & 1011(48)  & 1153(73)  \\
 &  &  & 0.01 & 467(26)  & 772(104)  & 1070(53)  & 1040(60)  & 1093(24)  & 1160(29)  \\
\cline{2-10}
 & $32^3 \times 12$ & 220 & 0.001 & 433(56)  & 445(55)  & 1047(36)  & 1048(30)  & 1066(42)  & 1060(43)  \\
 &  &  & 0.0025 & 486(25)  & 538(46)  & 1031(48)  & 1030(36)  & 1122(44)  & 1156(43)  \\
 &  &  & 0.00375 & 402(28)  & 698(269)  & 978(65)  & 1006(63)  & 1024(52)  & 942(51)  \\
 &  &  & 0.005 & 403(46)  & & 1054(47)  & 1073(32)  & 1252(94)  & 1094(108)  \\
 &  &  & 0.01 & 408(24)  & & 967(51)  & 1062(39)  & 1128(52)  & 1096(59)  \\
\cline{2-10}
 & $40^3 \times 12$ & 220 & 0.005 & 334(85)  & & 1022(37)  & 1021(37)  & 1068(69)  & 1146(45)  \\
 &  &  & 0.01 & 375(35)  & & 1000(21)  & 1040(33)  & 1056(55)  & 1228(77)  \\
\cline{2-10}
 & $48^3 \times 12$ & 220 & 0.001 & 569(35)  & 571(36)  & 1001(31)  & 1001(31)  & 1126(21)  & 1125(21)  \\
 &  &  & 0.0025 & 608(21)  & 616(21)  & 1002(34)  & 1003(34)  & 1114(40)  & 1113(40)  \\
 &  &  & 0.00375 & 577(66)  & 540(154)  & 944(51)  & 946(53)  & 992(66)  & 1045(76)  \\
 &  &  & 0.005 & 425(62)  & 581(125)  & 1085(28)  & 1092(28)  & 1091(59)  & 1122(68)  \\
\cline{2-10}
 & $32^3 \times 10$ & 260 & 0.005 & 959(18)  & 998(20)  & 1307(7)  & 1308(7)  & 1368(7)  & 1366(7)  \\
 &  &  & 0.008 & 941(19)  & 966(22)  & 1332(11)  & 1333(11)  & 1386(10)  & 1385(10)  \\
 &  &  & 0.01 & 850(25)  & 997(69)  & 1313(11)  & 1314(11)  & 1357(15)  & 1363(14)  \\
 &  &  & 0.015 & 935(50)  & 1144(90)  & 1381(13)  & 1387(14)  & 1426(14)  & 1436(15)  \\
\cline{2-10}
 & $32^3 \times 8$ & 330 & 0.001 & 1486(34)  & 1486(34)  & 1781(14)  & 1781(14)  & 1837(14)  & 1837(14)  \\
 &  &  & 0.005 & 1535(13)  & 1535(13)  & 1799(9)  & 1799(9)  & 1849(11)  & 1849(11)  \\
 &  &  & 0.01 & 1551(12)  & 1553(12)  & 1818(10)  & 1819(10)  & 1857(9)  & 1857(9)  \\
 &  &  & 0.015 & 1526(11)  & 1528(11)  & 1792(10)  & 1793(10)  & 1853(11)  & 1854(11)  \\
 &  &  & 0.02 & 1489(49)  & 1610(56)  & 1781(9)  & 1783(9)  & 1806(10)  & 1809(10)  \\
 &  &  & 0.04 & 1560(19)  & 1574(20)  & 1833(12)  & 1838(12)  & 1864(11)  & 1870(11)  \\
\cline{2-10}
\hline\hline
  \end{tabular}
  \caption{Meson screening mass determined with two-quark-inspired fit ansatz.
  Fit range is several lattice spacings around $z=8$ (0.6 fm) depending on parameters. }
  \label{tab:screeningmasses_mesons}
\end{table}


\subsection{Baryon correlators}

Finally, let us discuss baryon correlators.
We calculate the spatial correlation functions of baryon operators
projected onto positive $z$-parity (or an even component under the $z$ reflection)
and the lowest Matsubara frequency $\omega_0 = \pi T$,
which takes account of the anti-periodic boundary conditions in $t$-direction~\cite{DeTar:1987xb}:
\begin{eqnarray}
  C_j = \sum_{x,y,t} e^{it\omega_0} \langle N^+_j(x,y,z,t) \bar{N}^+_j(0,0,0,0)\rangle,
\end{eqnarray}
with nucleon operators $N^+_j=\hat{P}^z_+ (q^T i\tau_2\Gamma_j^1 q) \Gamma_j^2 q$,
and the parity projection operator $\hat{P}^z_+ = (1+\gamma_3)/2$.
The combinations $(\Gamma_j^1, \Gamma_j^2)$ specify the operator channels:
$N_1 = (C\gamma_5, 1), N_2 = (C, \gamma_5), N_3 = (C\gamma_4\gamma_5, 1)$,
and $N_4 = (C\gamma_4\vec\tau, \gamma_5\vec\tau)$ with the charge conjugation matrix $C=i\gamma_2\gamma_4$.
Similar to the case of mesons,
$N_1$ and $N_2$ channels are related by the axial $U(1)$ transformation,
and the $N_3$ -- $N_4$ pair probes $SU(2)_L\times SU(2)_R$ symmetry~\cite{Nagata:2007di}.

Along the lines of the two-quark-inspired function for mesons, we use a
three-quark-inspired function
\begin{eqnarray}
  \label{eq:3qfunc}
  B_j\frac{e^{-m'_jz}}{m'_jz^2}
\end{eqnarray}
to extract a screening mass $m'_j$ for each channel
by fitting the
forward propagating states.
This procedure gives qualitatively the same picture as in the previous section:
a more stable plateau located at a lower energy value
than that from the single $\cosh$ function.
We therefore use $m'$ to
compare masses of different channels.
Contrary to the case of mesons, we measure the baryon correlation functions
exclusively in $z$-direction without low-mode averaging.

\begin{figure*}[tbh]
  \centering
  \includegraphics[width=14cm]{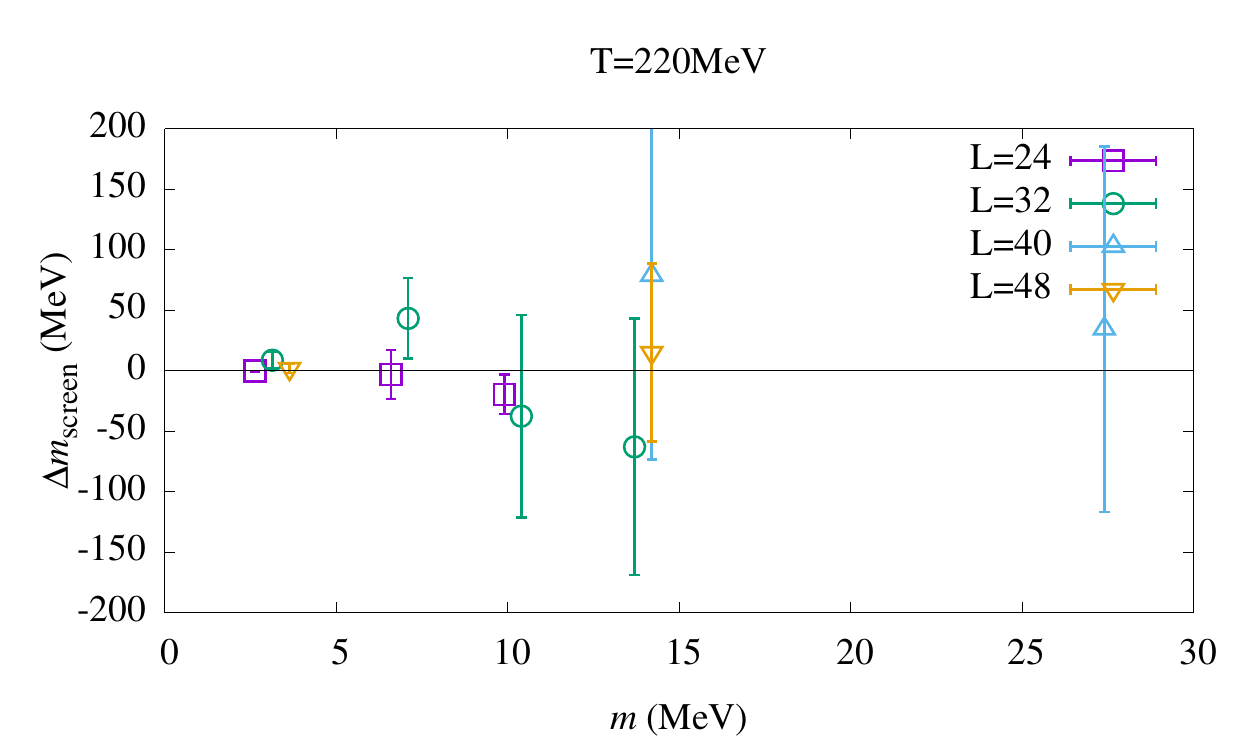}
    \caption{
      Difference of the fitted $m'$ between the baryon $N_1$ and $N_2$ correlators,
      which are connected by the $U(1)_A$ rotation.
      The data at $T=220$ MeV at different volumes are shown.
    }
    \label{fig:BaryondifU1}
  \includegraphics[width=14cm]{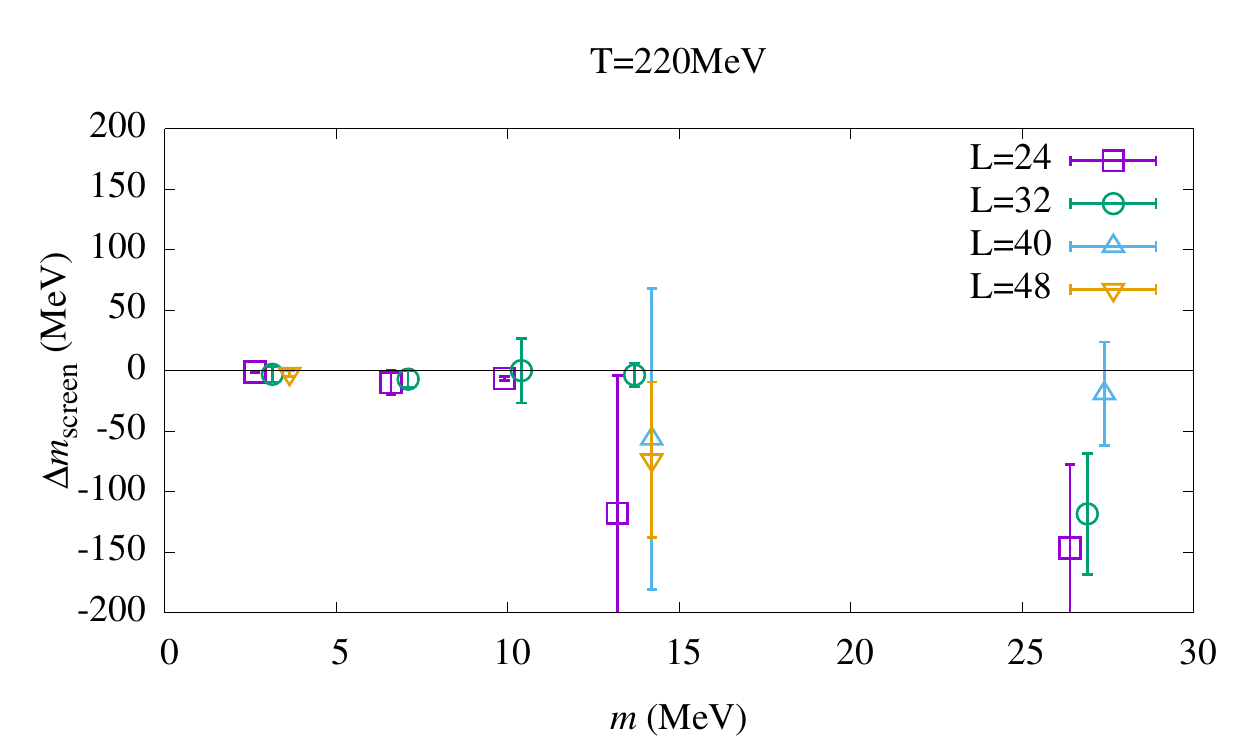}
  \caption{
    The same as Fig.~\ref{fig:BaryondifU1} but between the baryon $N_3$ and $N_4$ correlators,
    which shows the $SU(2)_L\times SU(2)_R$ symmetry.
  }
  \label{fig:BaryondifSU2}
\end{figure*}

The numerical results presented in Table \ref{tab:screeningmasses_baryons} and in Fig.~\ref{fig:BaryondifU1}
show the difference of $m'$ between the axial $U(1)$ partners.
While the uncertainty grows with increasing quark mass,
the signal is consistent with zero for all volumes at $T=$220 MeV.
A similar restoration pattern is seen for $SU(2)_L \times SU(2)_R$
symmetry, as shown in Fig.~\ref{fig:BaryondifSU2},
albeit with less fluctuations.

\begin{figure*}[tbh]
  \centering
  \includegraphics[width=8cm]{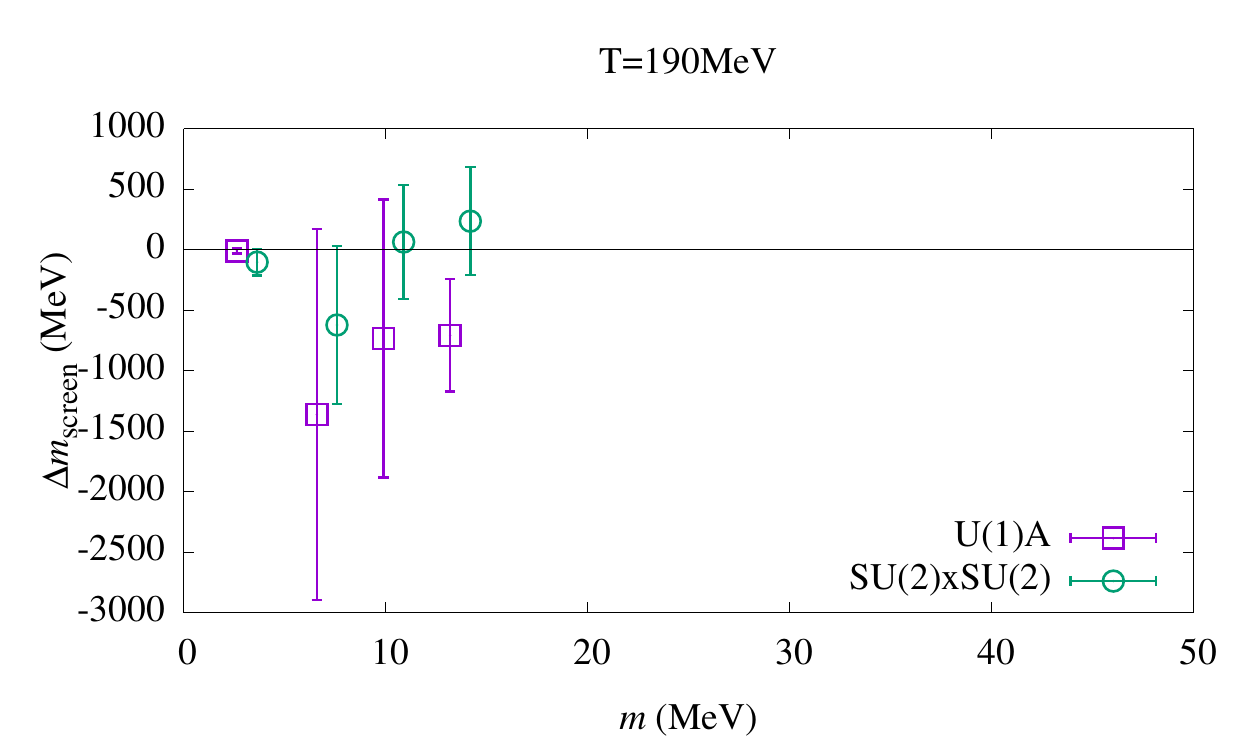}
  \includegraphics[width=8cm]{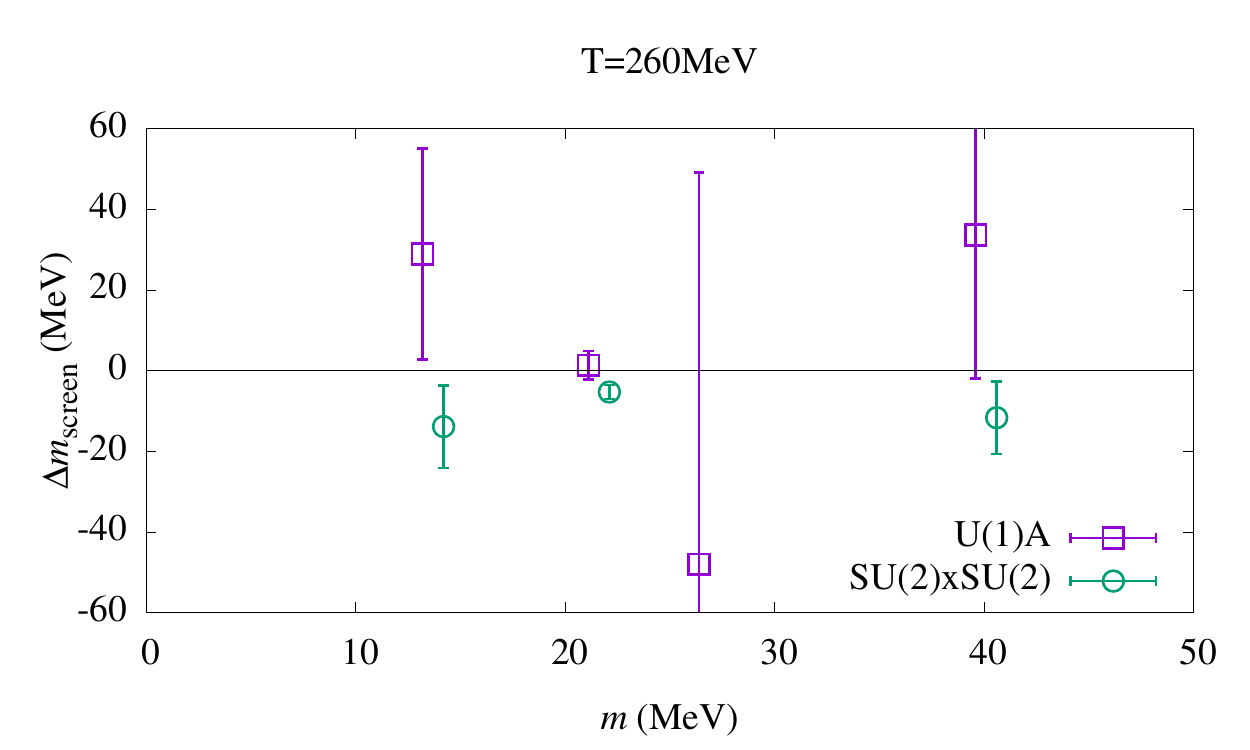}
  \includegraphics[width=8cm]{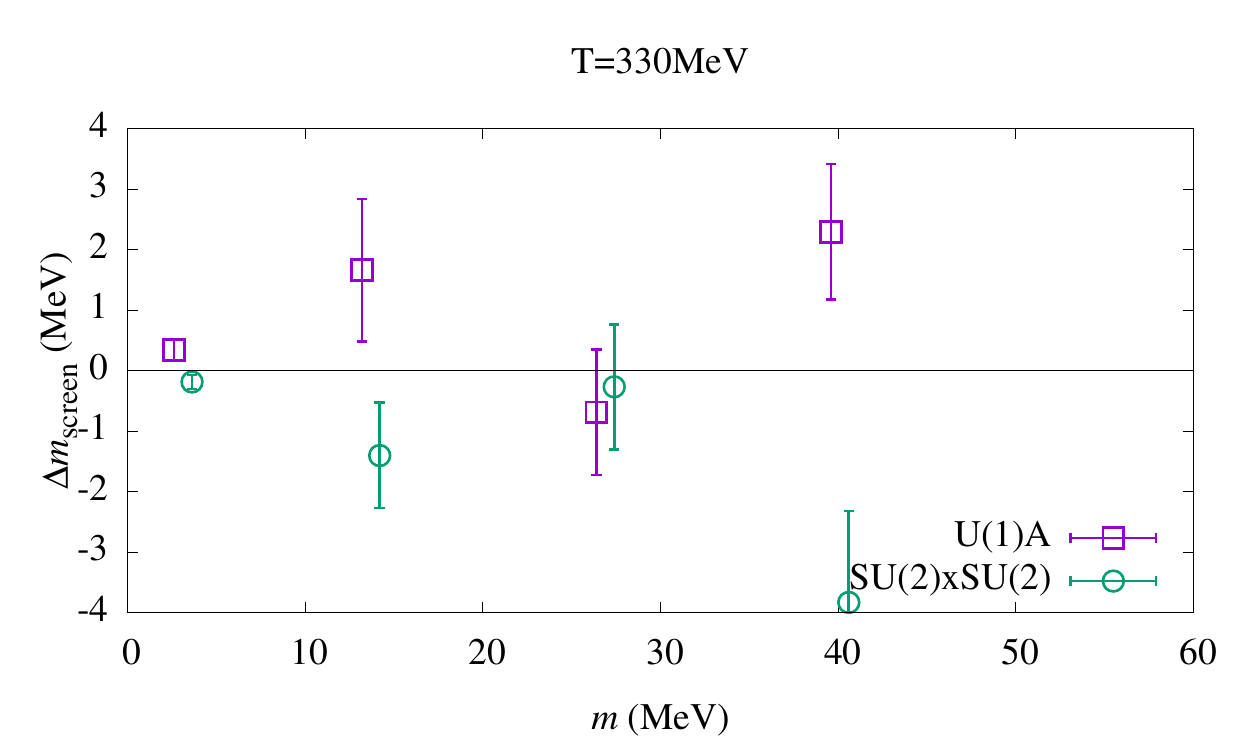}
  \caption{
    Difference of the fitted $m'$ between the baryon $N_1$ and $N_2$ correlators (squares)
    and that between the baryon $N_3$ and $N_4$ correlators (circles).
  }
  \label{fig:BaryondifT}
\end{figure*}

In Fig.~\ref{fig:BaryondifT} the mass difference for pairs of both
symmetries is shown at different temperatures.
At $T=190$ MeV, closer to the chiral transition, noise dominates all
quark masses except the lightest one. All data indicate consistency
with zero in the chiral limit.
At $T=260$ MeV and $T=330$ MeV some tiny violation at the order of
$\mathcal{O}(10)$ MeV can be seen for
non-vanishing quark masses. Similar to the meson screening masses,
this is $\sim 1\%$ of the individual screening masses $m'$.

\renewcommand{\arraystretch}{0.5}
\begin{table}[tbh]
  \centering
  \begin{tabular}{cccc|cccc}
\hline\hline
    &  &  &  & \multicolumn{4}{c}{$m'$(MeV)}\\
    $\beta$ & size & $T$(MeV) & $ma$ & $N_1$ & $N_2$ & $N_3$ & $N_4$  \\
\hline
4.30 & $32^3\times 14$ & 190 & 0.001   &    628(227) & 619(235) & 723(245) & 621(335) \\ 
                        &  &  & 0.0025 &      &   & 1882(409) & 1260(852) \\ 
                       &  &  & 0.00375 &      &   & 1576(403) & 1640(292) \\ 
                         &  &  & 0.005 &      &   & 1173(348) & 1410(150) \\ 
\cline{2-8}
     & $24^3\times 12$ & 220 & 0.001   &    1525(83) & 1525(83) & 1848(48) & 1847(48) \\ 
                       &  &  & 0.0025  &    1626(39) & 1623(41) & 1847(37) & 1838(45) \\ 
                       &  &  & 0.00375 &    1665(45) & 1645(46) & 1834(35) & 1828(36) \\ 
                       &  &  & 0.005   &    1351(100) & 2291(500) & 1863(72) & 1745(79) \\ 
                       &  &  & 0.01    &    1350(77) & 1949(200) & 1865(60) & 1719(63) \\ 
\cline{2-8}
    &  $32^3\times 12$ & 220 & 0.001   &    1563(49) & 1572(49) & 1700(35) & 1697(36) \\ 
                       &  &  & 0.0025  &    1440(48) & 1484(60) & 1647(49) & 1640(49) \\ 
                       &  &  & 0.00375 &    1507(81) & 1469(103) & 1687(41) & 1687(39) \\ 
                       &  &  & 0.005   &    1557(56) & 1494(97) & 1754(48) & 1750(46) \\ 
                       &  &  & 0.01    &    1343(95) & 1854(160) & 1763(50) & 1645(50) \\ 
\cline{2-8}
      &  $40^3\times 12$ & 220 & 0.005 &    1442(82) & 1521(116) & 1681(101) & 1624(70) \\ 
                         &  &  & 0.01  &    1425(93) & 1459(166) & 1692(52) & 1673(44) \\ 
\cline{2-8}
    &  $48^3\times 12$ & 220 & 0.001   &    1416(122) & 1418(123) & 1307(113) & 1305(114) \\ 
                       &  &  & 0.0025  &       &    & 1466(295) & 1999(266) \\ 
                       &  &  & 0.00375 &       &     & 1773(234) & 1403(196) \\ 
                       &  &  & 0.005   &    1605(56) & 1620(50) & 1716(59) & 1643(52) \\ 
\cline{2-8}
    &  $32^3\times 10$ & 260 & 0.005   &    2042(23) & 2071(25) & 2163(19) & 2149(20) \\ 
                         &  &  & 0.008 &    2018(32) & 2020(32) & 2160(29) & 2154(29) \\ 
                         &  &  & 0.01  &    1950(91) & 1902(131) & 2226(64) & 2061(77) \\ 
                         &  &  & 0.015 &    2080(44) & 2114(53) & 2207(43) & 2196(40) \\ 
\cline{2-8}
      & $32^3\times 8$ & 330 & 0.001   &    2847(34) & 2847(34) & 2948(30) & 2948(30) \\ 
                         &  &  & 0.005 &    2727(38) & 2728(38) & 2835(39) & 2833(39) \\ 
                         &  &  & 0.01  &    2775(45) & 2775(45) & 2900(50) & 2899(49) \\ 
                         &  &  & 0.015 &    2773(19) & 2776(19) & 2873(17) & 2869(17) \\ 
                         &  &  & 0.02  &    2740(32) & 2745(32) & 2864(30) & 2859(30) \\ 
                         &  &  & 0.04  &    2710(28) & 2733(30) & 2857(23) & 2841(22) \\ 
\hline\hline
  \end{tabular}
  \caption{Baryon screening masses determined with the three-quark-inspired fit ansatz.}
  \label{tab:screeningmasses_baryons}
\end{table}

\clearpage
\section{Conclusion}
\label{eq:conclusion}

In this work, we simulated two-flavor lattice QCD
and tried to quantify how much of the axial $U(1)$ anomaly survives
at high temperatures $190$--330 MeV.
We employed the M\"obius domain-wall fermion action
and the overlap fermion action whose determinant
is obtained by a stochastic reweighting technique.
We fixed the lattice spacing to 0.074 fm, and
chose more than four quark masses, including
one below the physical point.

We confirmed that our data are consistent with
those in the previous works \cite{Tomiya:2016jwr}
extending statistics of ensembles at $\beta=4.24$.
We also observed a good consistency between
the M\"obius domain-wall and overlap fermions,
except for the axial $U(1)$ susceptibility, which
is very sensitive to the violation of the chiral symmetry at $T=220$ MeV.
The discretization effect is therefore well under control.
We also confirmed that the systematics due
to finite size of the lattice is under control.
Our data with various lattice sizes agree, except for
those with $L=24$, which has a small aspect ratio $TL=2$ at $T=220$ MeV.

In the Dirac spectrum we found a strong suppression
of low but non-zero eigenmodes.
The higher temperature,
the more suppression of the low lying modes observed.
On the other hand, for the chiral zero mode,
a peak is found at all four
simulated temperatures but 
its quark mass dependence is steep and
the chiral limit is consistent with zero.

As expected from the behavior of the chiral zero mode,
a sharp disappearance of the topological susceptibility is found,
which suggests a mass dependence starting with a power $\sim m^4$
near the chiral limit.
Our numerical data for the axial U(1) susceptibility,
meson and baryon correlators also indicate that the
axial $U(1)$ anomaly is consistent with zero in the chiral limit.
From these observations we conclude that
the remaining anomaly of the axial $U(1)$ symmetry at the physical point
for $T\geq$ 1.1 $T_{c}$
is at most a few MeV level, which is $\sim 1$\% of the simulated temperatures.

To examine if the disappearance of the $U(1)_A$ anomaly occurs at the same time
as the $SU(2)_L\times SU(2)_R$ symmetry is restored,
  we need a simulation around
  the critical temperature, which is beyond the scope of this paper.

\begin{acknowledgments}
We thank T. Cohen, H.-T. Ding, C. Gattringer, L. Glozman, A. Hasenfratz, C.B. Lang, R. Pisarski, S. Prelovsek, for
useful discussions. We thank P. Boyle for correspondence for starting simulation with Grid on Intel Knights Landing machines,
and I. Kanamori for helping us on the simulations on K computer.
We also thank the members of JLQCD collaboration for their encouragement and support.
Numerical simulations are performed on IBM System Blue Gene Solution at KEK under a
support of its Large Scale Simulation Program (No. 16/17-14) and Oakforest-PACS at JCAHPC
under a support of the HPCI System Research Projects (Project IDs: hp170061, hp180061,
hp190090, and hp200086), Multidisciplinary Cooperative Research Program in CCS, University of Tsukuba
(Project IDs: xg17i032 and xg18i023) and K computer provided by the RIKEN Center for Computational Science.
We used Japan Lattice Data Grid (JLDG) for storing a part of
    the numerical data generated for this work.
This work is supported in part by the Japanese Grant-inAid for Scientific Research
(No. JP26247043, JP16H03978, JP18H01216, JP18H04484, JP18H05236), and by MEXT as
“Priority Issue on Post-K computer" (Elucidation of the Fundamental Laws and Evolution of the
Universe) and by Joint Institute for Computational Fundamental Science (JICFuS).
\end{acknowledgments}


\appendix
\section{Spectral function of two and three non-interacting quarks}
\label{app:spec}

In this appendix, we compute propagators for two and three non-interacting quarks
in $d$-dimensions (to show that $d=4$ is special).
To take the finite temperature into account,
the spacetime is assumed to be an Euclidean flat continuum
space with one direction compactified.
Namely, we consider $S^1\times \mathbb{R}^{d-1}$, and
anti-periodic boundary conditions are imposed on the fermions.
We denote the compact direction by $x_0$ and
consider spatial propagators in the $x_1$ direction.

\subsection{Two quarks}

Let us start with the non-interacting pseudoscalar ``meson'' propagator,
which is expressed by two massless and non-interacting quarks.
By the standard Fourier transformation
we obtain
\begin{eqnarray}
  C_{2q}(x_1)&\equiv& \int_{S^1\times \mathbb{R}^{d-2}} d^{d-1}x \langle \bar{d}\gamma_5 u(x)\bar{u}\gamma_5 d(0)\rangle
  \nonumber\\&=&
  \int_{S^1\times \mathbb{R}^{d-2}} d^{d-1}x {\rm tr}[\gamma_5 D^{-1}(x,0)\gamma_5 D^{-1}(0,x)]
  \nonumber\\&=&
  4\int_{S^1\times \mathbb{R}^{d-2}} d^{d-1}x \int \frac{d^d p}{(2\pi)^d}\int \frac{d^d p'}{(2\pi)^d}
  \frac{p^\mu p_\mu'}{(p)^2(p')^2}e^{i(p-p')_\nu x^\nu}
  \nonumber\\&=&
   4\int \frac{d^{d-1} \bm{p}}{(2\pi)^{d-1}}\int \frac{dp_1}{2\pi}\int \frac{dp'_1}{2\pi}\frac{(p_1p'_1+\bm{p}^2)}{(p_1^2+\bm{p}^2)(p_1^{\prime 2}+\bm{p}^2)}e^{i(p_1-p_1')x_1}.
\end{eqnarray}
Noting that the 0-th component of $\bm{p}$ denoted by $p_0$ is discrete,
and neglecting
higher $p_0$ contribution except for the lowest Matsubara frequency $p_0=M=\pm \pi T$,
we can use the following approximation
\begin{eqnarray}
   \int \frac{d^{d-1} \bm{p}}{(2\pi)^{d-1}} \sim 2T \int \frac{d^{d-2} \bm{q}}{(2\pi)^{d-2}},\;\;\; \bm{p}^2 = M^2+\bm{q}^2,
\end{eqnarray}
where the $(d-2)$-dimensional vector $\bm{q}$ is given by $\bm{q}=(p_2,p_3,\cdots p_{d-1})$ and the factor two
comes from the two possible signs of $M$.
Changing the variables as $P_1=p_1-p'_1$, $R_1=(p_1+p'_1)/2$ and explicitly integrating over $R_1$ and $P_1$,
we obtain
\begin{eqnarray}
  C_{2q}(x_1) &=& 4T \int \frac{d^{d-2} \bm{q}}{(2\pi)^{d-2}} e^{-2\sqrt{M^2+\bm{q}^2}x_1} = 4CT \int dq q^{d-3}e^{-2\sqrt{M^2+q^2}x_1}
  \nonumber\\&=& CT\int^{\infty}_{2M} d\omega \omega \left(\sqrt{\frac{\omega^2}{4}-M^2}\right)^{d-4} e^{-\omega x_1},
\end{eqnarray}
where the constant $C$ comes from the solid angle integral.
In the last line we have changed the integral variable to
$\omega=2\sqrt{M^2+q^2}$.
Note that a fractional power is absent for $d=4$.

From the above integral, we can read off the spectral function for $d=4$ as
\begin{eqnarray}
\rho_{2q}^{\rm free}(\omega)=2CT\theta(\omega-2M)\left[2M+(\omega-2M)\right],
\end{eqnarray}
which supports our assumption for the fitting form Eq.~(\ref{eq:2qfunc}) of the meson correlators.
Here we have chosen the pseudoscalar correlators but
it was confirmed in \cite{Rohrhofer:2017grg} by a full computation including higher Matsubara frequencies
that this asymptotic form is universal in all other channels.

\subsection{Three quarks}

Next let us consider a ``baryon'' two-point function
in which three non-interacting quarks propagate, choosing
the $N_1$ channel,
\begin{eqnarray}
   \int_{S^1\times \mathbb{R}^{d-2}} d^{d-1}x e^{iMx_0}\langle [(u^TC\gamma_5d) u](x)[\bar{u}(\bar{d}C\gamma_5\bar{u}^T](0)\rangle.
\end{eqnarray}
One of the contractions leads to
\begin{eqnarray}
  C_{3q}(x_1)&\equiv&\int_{S^1\times \mathbb{R}^{d-2}} d^{d-1}x e^{iMx_0} D^{-1}(x,0){\rm tr}[\gamma_5 D^{-1}(x,0)\gamma_5 D^{-1}(0,x)]
  \nonumber\\&=&
  4\int_{S^1\times \mathbb{R}^{d-2}} d^{d-1}x \int \frac{d^d p}{(2\pi)^d}\int \frac{d^d p'}{(2\pi)^d}\int \frac{d^d p''}{(2\pi)^d}
  \frac{(p_\mu p^\mu)(p''_\nu\gamma^\nu)}{(p)^2(p')^2(p'')^2}e^{i(p-p'-p'')_\nu x^\nu+iMx_0}
  \nonumber\\&=&
   4\int \frac{d^d p}{(2\pi)^d}\int \frac{d^d p'}{(2\pi)^d}\int \frac{d p''_1}{(2\pi)}
  \frac{(p_\mu p^\mu)(p''_\nu\gamma^\nu)}{(p)^2(p')^2(p'')^2}e^{i(p-p'-p'')_1 x_1},
\end{eqnarray}
where $p''_\mu=(p_0-p'_0-M, p''_1, \bm{q}-\bm{q}')$ with  $(d-2)$-momentum vectors
$\bm{q}=(p_2,p_3,\cdots, p_{d-1})$ and $\bm{q}'=(p'_2,p'_3,\cdots, p'_{d-1})$.
In the same way as the two-quark propagation, let us ignore the summation over $p_0, p_0'$
except for the three cases with $p_0-p_0'-M=\pm M$.
We then obtain
\begin{eqnarray}
  C_{3q}(x_1)&\sim&
  12T^2\int \frac{d^{d-2} \bm{q}}{(2\pi)^{d-2}}
  \int \frac{d^{d-2} \bm{q}'}{(2\pi)^{d-2}}
  \int \frac{d p_1}{(2\pi)}
   \int \frac{d p'_1}{(2\pi)}
   \int \frac{d p''_1}{(2\pi)}
   \nonumber\\&&
   \frac{(p_1p'_1+M^2+\bm{q}\cdot \bm{q}')(p''_1\gamma_1+M\gamma_0+(\bm{q}-\bm{q}')\cdot \bm{\gamma})}{(p_1^2+M^2+\bm{q}^2)(p_1^{\prime 2}+M^2+\bm{q'}^2)(p_1^{''2}+M^2+(\bm{q}-\bm{q}')^2)}e^{i(p-p'-p'')_1 x_1}
   \nonumber\\&=&
   \frac{3T^2}{2}\int \frac{d^{d-2} \bm{q}}{(2\pi)^{d-2}}
  \int \frac{d^{d-2} \bm{q}'}{(2\pi)^{d-2}}e^{-(\sqrt{M^2+\bm{q}^2}+\sqrt{M^2+\bm{q'}^2}+\sqrt{M^2+(\bm{q}-\bm{q}')^2}) x_1}
   \nonumber\\&&\times
  {\textstyle
   \frac{(\sqrt{M^2+\bm{q}^2}\sqrt{p_1^{\prime 2}+M^2+\bm{q'}^2}+M^2+\bm{q}\cdot \bm{q}')
     (\sqrt{p_1^{''2}+M^2+(\bm{q}-\bm{q}')^2}\gamma_1+M\gamma_0+(\bm{q}-\bm{q}')\cdot \bm{\gamma})}
        {\sqrt{M^2+\bm{q}^2}\sqrt{p_1^{\prime 2}+M^2+\bm{q'}^2}\sqrt{p_1^{''2}+M^2+(\bm{q}-\bm{q}')^2}}
        }.       
\end{eqnarray}
As large $\bm{q}^2, \bm{q}^{\prime 2}$ contributions are exponentially suppressed,
let us expand the integrand with $\bm{q}^2/M$ and $\bm{q'}^2/M$ so that the
integral is greatly simplified as
\begin{eqnarray}
  C_{3q}(x_1)&\sim&
   3T^2\int \frac{d^{d-2} \bm{q}}{(2\pi)^{d-2}}
   \int \frac{d^{d-2} \bm{q}'}{(2\pi)^{d-2}}(\gamma_1+\gamma_0)
   e^{-x_1(3M+(\bm{q}^2+\bm{q'}^2-\bm{q}\cdot \bm{q'})/M)}
   \nonumber\\&\propto &
   T^2 \int_0^\infty dp p^{d-3}\left(\frac{M}{x_1}\right)^{(d-2)/2}e^{-3x_1(M+p^2/4M)}
   \nonumber\\&\propto &
   T^2 \int_{3M}^\infty d\omega M^{d-2}(\omega-3M)^{\frac{d-4}{2}}x_1^{-\frac{d-2}{2}}e^{-\omega x_1}
    \nonumber\\&\propto &
   T^2 \int_{3M}^\infty d\omega M^{d-2}(\omega-3M)^{d-3}e^{-\omega x_1},
\end{eqnarray}
where the integration over $\bm{q}'-\bm{p}/2$ and solid angle in $\bm{q}$
are performed and an unimportant overall dimensionless (matrix-valued) constant  is neglected.

From the above integral, we can read off the spectral function for $d=4$ as
\begin{eqnarray}
\rho_{3q}^{\rm free}(\omega)=DT^4\theta(\omega-3M)(\omega-3M),
\end{eqnarray}
with a constant (matrix) $D$,
which supports the asymptotic three-quark form $\exp(-3Mx_1)/x_1^2$
corresponding to Eq.~(\ref{eq:3qfunc}).

\end{document}